\documentclass[12pt]{article}
\usepackage{amsmath,amssymb,epsfig}
\usepackage{cancel}
\usepackage{color}

\makeatother
\newcommand{\III}{I\hspace{-.1em}I\hspace{-.1em}I}

\begin{document}

\title{
\vbox{
\baselineskip 14pt 
\hfill \hbox{
\normalsize KEK-TH-1924, KUNS-2637}\\
} \vskip 1.7cm
\bf Reheating era leptogenesis\\ in models with a seesaw mechanism
\vskip 0.5cm}
\author{
Yuta~Hamada$^1$,~
Koji~Tsumura$^2$,~
Daiki~Yasuhara$^2$
\\*[20pt]
$^1${\it \normalsize 
KEK Theory Center, IPNS, KEK, Tsukuba, Ibaraki 305-0801, Japan }\\
%\\*[50pt]
$^2${\it \normalsize 
Department of Physics, Kyoto University, Kyoto 606-8502, Japan }\\
\\*[50pt]}

\date{
\centerline{\small \bf Abstract}
\begin{minipage}{0.9\linewidth}
\medskip 
\medskip 
\small
Observed baryon asymmetry can be achieved not only by the decay of right-handed neutrinos 
but also by the scattering processes in the reheating era. % \cite{Hamada:2015xva}. 
In the latter scenario, new physics in high energy scale does not need to be specified, 
but only two types of  the higher dimensional operator of 
the standard model particles are assumed in the previous work. 
In this paper, we examine the origin of the higher dimensional operators assuming 
models with a certain seesaw mechanism at the high energy scale. 
The seesaw mechanism seems to be a simple realization of the reheating era leptogenesis 
because the lepton number violating interaction 
is included. 
We show that the effective interaction giving CP violating phases is provided in the several types of models 
and also the reheating era leptogenesis actually works in such models. 
Additionally, we discuss a possibility for lowering the reheating temperature in the radiative seesaw models, 
where the large Yukawa coupling is naturally realized.  
\end{minipage}
}

\newpage

\begin{titlepage}
\maketitle
\thispagestyle{empty}
\clearpage
\end{titlepage}

\renewcommand{\thefootnote}{\arabic{footnote}}
\setcounter{footnote}{0}

%============================================================================

\section{Introduction}
\label{sec1}

The standard model (SM) for elementary particles serves as the most reliable framework 
to explain observed phenomena in particle physics so far. 
Since no signature of new physics beyond the SM is found at the TeV scale, 
some people start to consider seriously the possibility that 
the minimal SM works up to the very high energy scale. 
In fact, the observed value of the Higgs boson mass 
not only suggests the Higgs coupling to be perturbative up to high energy 
but also implies a critical behavior at around the Planck scale, see Ref.~\cite{Hamada:2012bp} for example. 
On the other hand, it is true that many problems such as baryon asymmetry of the universe, 
the origin of neutrino mass, existence of the cosmic dark matter are left unsolved in the SM. 

The observed value of the baryon asymmetry is~\cite{Ade:2015xua}, 
\begin{align}
\frac{n_B^{}}{s} 
\simeq 
(8.67\pm0.05)\times10^{-11}, 
\label{eq:obseved}
\end{align}
where $n_B^{}$ is the baryon number density and $s$ is the entropy density. 
Although the SM satisfies Sakharov's three conditions for the baryogenesis, 
the SM cannot accommodate a sufficient amount of the baryonic matter in the universe 
because of the smallness of the violation of the CP symmetry and the lack of the first order phase transition at the electroweak scale.
In models of physics beyond the SM, many baryogenesis scenarios have been suggested.\footnote{
See Ref.~\cite{decay} for earlier discussion of baryogenesis via delayed decay of heavy particles.  
} 
Well known examples include the GUT baryogenesis~\cite{Yoshimura:1978ex}, 
leptogenesis~\cite{Fukugita:1986hr}, 
Affleck-Dine baryogenesis~\cite{Affleck:1984fy}, 
electroweak baryogenesis~\cite{Kuzmin:1985mm} and 
string scale baryogenesis~\cite{Aoki:1997vb}, etc. 

The leptogenesis would be the most simple scenario, 
where only the singlet right-handed neutrinos are added to the SM. 
In this scenario, the smallness of the left-handed neutrinos is explained by 
the super-heavy right-handed neutrinos through the Type-I seesaw mechanism~\cite{typeI}. 
At the same time, the lepton number asymmetry is created by the decay of heavy right-handed neutrinos, 
and is converted into that of the baryon number via the sphaleron process~\cite{'tHooft:1976fv}. 
It is quite economical scenario in a sense that the lepton number is naturally violated 
by the Majorana mass term of the right-handed neutrinos, and 
the out-of-equilibrium condition is satisfied by the decay of heavy particles.\footnote{Right-handed neutrinos are considered to be produced thermally 
or by the decay of an inflaton in the early universe~\cite{Lazarides:1991wu}. } 

Recently, another way to achieve the leptogenesis scenario is suggested in Ref.~\cite{Hamada:2015xva}. 
We here call it the reheating era leptogenesis, while the original one is called the conventional leptogenesis. 
In this new scenario, the lepton number asymmetry is generated by the scattering of the SM particles, 
while the out-of-equilibrium is realized since the high energy SM particles are provided by the decay of 
the assumed inflaton at the reheating era. 
The heavy particles other than the inflaton are not necessarily produced at on-shell. 
Instead, only the effective (higher dimensional) interactions among the SM particles 
for the scattering processes and for the CP violation are introduced 
to describe the reheating era leptogenesis. 
Thus, the detailed structure of the new physics model at the high energy scale 
does not need to be specified. 

As the underlying theory of such interactions, 
many variants of neutrino mass generation models can be considered as a candidate. 
There are three types of the seesaw mechanism at the tree-level, 
where the dimension-five operator for the origin of the left-handed Majorana neutrino masses 
is decomposed only by the single particle. 
The Type-I (-\III)~\cite{typeI,typeIII} seesaw mechanism introduces SU(2)${}_L$ singlet (triplet) fermions, 
on the other hand, the Type-II~\cite{typeII} does a triplet scalar field with a vacuum expectation value (VEV). 
If we add more than or equal to two kinds of particles, 
the neutrino masses can be generated by the quantum loop effect~\cite{Ref:Zee,Ref:Zee-Babu}. 
In this class of models, small neutrino masses are realized not only by heavy new particles 
but also by the loop suppression factor(s). 
Another advantage is that the new particle inside loop(s) can be identified 
as the dark matter in some models~\cite{Krauss:2002px}. 

In this paper, we extend the analysis of the letter article~\cite{Hamada:2015xva}.
We review the reheating era leptogenesis~\cite{Hamada:2015xva} and 
apply some variations of the seesaw mechanism to this scenario
as the concrete examples of new physics models at the high energy scale. 
An additional contribution from the lepton number violating collision, 
which is not considered in the letter paper~\cite{Hamada:2015xva}, is also taken into account. 
Various kinds of constraints such as upper bounds on the inflaton mass, 
a perturbativity bound on the Yukawa coupling, and 
constraints from efficiency factors are studied. 
Under these conditions, we show that the reheating era leptogenesis can be realized 
in the wide range of the parameter space in each model. 
We also derive the upper bound on the reheating temperature, which comes from the strong washout effect.
Furthermore, in a radiative seesaw model, the reheating temperature is lowered 
without introducing the fine-tuning among the parameters, 
because the Yukawa coupling can be much larger than that in the Type-I seesaw model.

This paper is organized as follows. %......
In Section \ref{sec2}, we review the reheating era leptogenesis, 
and summarize Boltzmann equations used in this paper.
In Section \ref{sec3}, 
the reheating era leptogenesis scenarios are discussed in 
models of the seesaw mechanism including not only the tree-level seesaw 
but also the radiative seesaw mechanisms. 
Section \ref{sec4} is devoted to conclusion and discussion.

%============================================================================

\section{The reheating era leptogenesis scenario} 
\label{sec2}

\begin{figure}[tb]
 \centering
 \includegraphics[height=3.5cm]{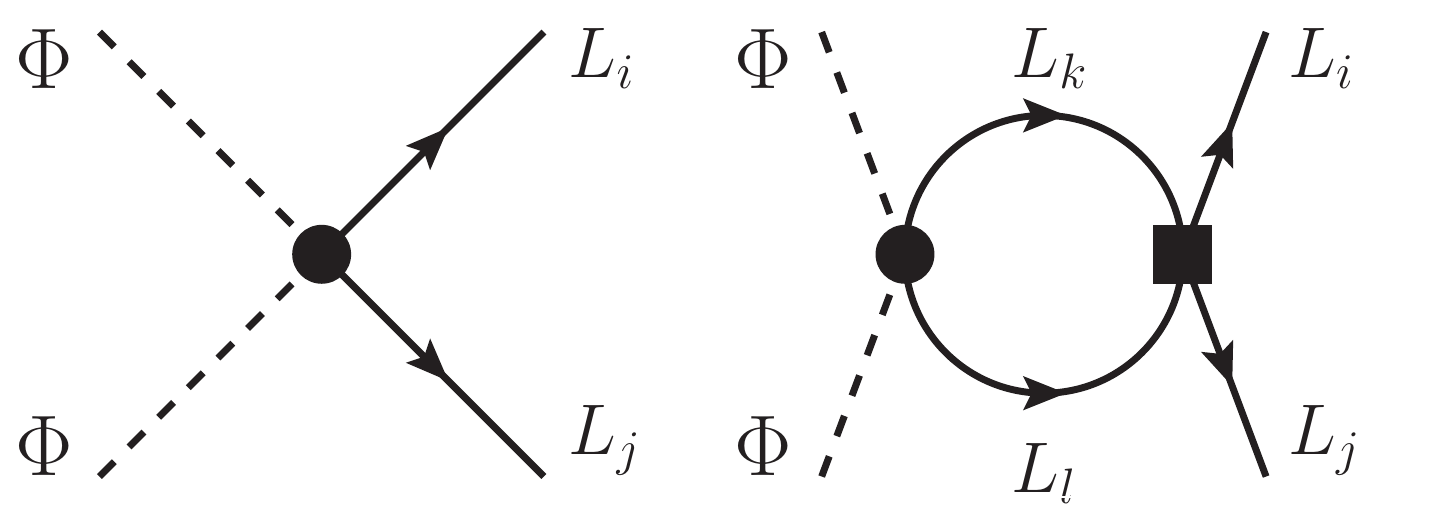}
  \caption{Interference between tree and one-loop diagram for the lepton number violation scattering process. }  \label{fig:leptogenesis}
\end{figure}

In the reheating era leptogenesis scenario~\cite{Hamada:2015xva}, in addition to the inflaton and the SM fields, 
only two effective interactions are assumed as 
\begin{align}
\Delta{\mathcal L} 
   =\frac{\lambda^{(1)}_{ij}}{\Lambda_1 }(\overline{L_i}\widetilde{\Phi}) (\overline{L_j}\widetilde{\Phi}) 
      +\frac{\lambda^{(2)}_{ijkl}}{\Lambda_2^2} (\overline{L_i} \gamma^\mu L_j) (\overline{L_k} \gamma_\mu L_l)
         +\text{H.c.},
\label{eq:leptogenesis}
\end{align}
where $L_i$ is the left-handed lepton doublet, and $\Phi$ is the Higgs doublet. 
The coefficients $\lambda^{(1)}_{ij}/\Lambda_1$ is determined by the generic seesaw relation; 
\begin{align}
m_{\nu, i}^{}
=
\frac{\lambda^{(1)}_{ii} \, v^2}{\Lambda_1}, 
\label{eq:seesaw}
\end{align}
where $m_{\nu, i}^{}$ is the $i$-th mass eigenvalue of active (left-handed) neutrinos, and 
the VEV of the Higgs doublet field is given by $\langle \Phi\rangle = (0, v/\sqrt2)^T$ with $v=(\sqrt2G_F)^{-1}$. 
When we specify the ultraviolet theory, $\lambda^{(2)}_{ijkl}/\Lambda_2^2$ can also be fixed. 
We here choose the real diagonal basis of the coupling matrix $\lambda^{(1)}_{ij}$ 
by the unitary transformation of the leptonic SU(2)${}_L$ doublet. 
In this basis, the Yukawa couplings for the charged leptons can have physical complex phases. 
A typical magnitude of $\Lambda_2$ derived from the charged lepton Yukawa couplings is 
$\Lambda_2 \simeq (4\pi) (v/m_\tau^{})^2\, \sqrt{M_\text{inf}^{}T_R} \sim 10^5 \times \sqrt{M_\text{inf}^{}T_R}$, 
where the inflaton mass is $M_\text{inf}^{}$, and the reheating temperature $T_R$ is defined by 
the temperature $T$ of the thermal plasma at the time when the expansion rate of the universe 
balances with the inflaton decay rate $\Gamma_\text{inf}$, that is, 
$T_R=\left({3\over5}{90\over\pi^2 g_*}\Gamma_\text{inf}^2M_\text{Pl}^2\right)^{1/4}$.
Here $g_*$ is the effective numbers of relativistic degrees of freedom, which is $106.75$ 
in the SM at the temperature higher than the electroweak scale, and $M_\text{Pl}$ is the reduced Planck scale. 
These contributions are expected to be much smaller than those from the new physics beyond the SM, 
so that we can safely neglect these contributions in the following discussions. 
The first term in Eq.~\eqref{eq:leptogenesis} violates the lepton number by two units 
after the electroweak symmetry breaking, but with only this term non-zero baryon asymmetry cannot be created. 
Complex phases for the CP violation appear in the second term  in Eq.\eqref{eq:leptogenesis}. 
The net lepton number is produced by the scattering process via the interference 
between the tree and one-loop diagrams in Fig.~\ref{fig:leptogenesis}, 
where both the lepton number violation and the CP violation effects are included. 
The dimension-five (-six) vertices are denoted by the circle (square) symbols. 
In the reheating era leptogenesis scenario, the lepton asymmetry is created during 
the thermalization process of the SM particle after the inflation. 
The left-handed leptons are produced by the direct decay of the inflaton, and
are thermalized through the scattering with the SM particles in thermal plasma. 
This thermalization process proceeds in the out-of-equilibrium. 
During this era, the lepton asymmetry is generated by the process in the Fig.~\ref{fig:leptogenesis}. 
The baryon asymmetry is obtained similarly to the conventional leptogenesis 
by the conversion through the sphaleron process.

The baryon asymmetry can be evaluated by solving the following Boltzmann equations 
numerically~\cite{Hamada:2015xva},\footnote{
Comparing with the Boltzmann equations in Ref.~\cite{Hamada:2015xva}, we add the $\epsilon_2$ term 
in the right hand side of the second equation. 
%\blue{The $\epsilon_2$ term plays an important role for large $M_\text{inf}$ region as we will see later.}
}
\begin{align}\label{Eq:Boltzmann1}
& \dot{\rho}_R^{}+4\,H\,\rho_R^{}
   = \Big(1-\sum_{i} \mathcal{B}_i \Big)\, \Gamma_\text{inf}\, \rho_\text{inf}^{} 
        +\frac{M_\text{inf}^{}}{2}\, \sum_i n_{\ell_i}^{}\, \Gamma_\text{brems}, \\ 
\label{Eq:Boltzmann2}
& \dot{n}_L^{}+3\,H\,n_L^{} 
   =4\,\sum_i\,\epsilon_i^{}\,\Gamma_{\cancel{L}_i}\, n_{\ell_i}^{} 
      +2\,\sum_i\,\epsilon_{2i}^{}\,\Gamma_{2\cancel{L}_i}\, n_{\ell_i}^{} 
      -\Gamma_\text{wash}\,n_L^{}, \\ 
\label{Eq:Boltzmann3}
& \dot{n}_{\ell_i}^{}+3\,H\,n_{\ell_i}^{}
   =\frac{\Gamma_\text{inf}\,\rho_\text{inf}^{}}{M_\text{inf}^{}}\, \mathcal{B}_i
   -n_{\ell_i}^{}\,(\Gamma_\text{brems}+H), 
\end{align}
where $i=1,2,3$, $\rho_R^{} = \pi^2g_*T^4/30$, $\rho_\text{inf}^{}=\Lambda_\text{}^4\, e^{-\Gamma_\text{inf}\,t}/a(t)^3$ 
are the energy densities of the radiation and the inflaton, respectively. 
$\mathcal{B}_i \equiv \mathcal{B}(\varphi\to L_iX)$ is the branching fraction of the inflaton $\varphi$  
into a $\overline{L_i}$ and other particles.\footnote{
The decays of the inflaton depend on the detailed models of the inflaton interaction. 
For instance, we may consider a dimension-five operator as 
$\varphi\,\overline{L_i}e_{R\ell}^{}\Phi$. 
If the minimal flavor violation hypothesis is imposed, the coupling matrix in our basis is
$(y_e^{})_{i\ell}^{} = \sqrt2\, U_{\ell i}^* M_\ell^\text{diag}/v$. 
Thus, branching ratios have the specific structure, i.e., 
$\mathcal{B}_i = \sum_\ell |(y_e^{})_{i\ell}|^2 / \sum_{j\ell} |(y_e^{})_{j\ell}|^2 \approx  |U_{\tau i}|^2,$ where $U_{f i}$ is the PMNS matrix~\cite{PMNS}.
If we additionally introduce a flavor universal interaction such as $\varphi\,\overline{L_i}\cancel{D}L_i$, 
which cannot generate the baryon asymmetry. 
Then, $\mathcal{B}_i$ is simply reduced by a factor.  
In our numerical analysis, we assume only the former dimension-five interaction for simplicity and concreteness. 
}
The height of the potential during the inflation is $\Lambda_\text{inf}^4$. 
The created asymmetry is not sensitive to the value of $\Lambda_\text{inf}$, 
which is taken to be $\Lambda_\text{inf}=10^{15}$GeV in this paper. 
The scale factor $a(t)$ of the universe is related to the Hubble parameter $H = \dot{a}(t)/a(t)$, which is given by 
\begin{align}
H^2 = 
\frac1{3M_\text{Pl}^2} \Big( \rho_\text{inf}^{} +\rho_R^{} +\frac{M_\text{inf}^{}}2 \sum_i n_{\ell_i}^{} \Big), 
\end{align}
The number density $n_\ell^{}$ of the left-handed leptons is produced by the inflaton decay. 
That of the lepton asymmetry is denoted by $n_L^{}$. 
The factors efficiency $\epsilon_i^{}$ and $\epsilon_{2i}^{}$ represent the interference effect 
between the tree and one-loop diagrams, 
\begin{align}
\epsilon_{(2)i}^{} = 2\, 
\frac{\sigma_{\overline{L}_i\overline{L}_i \to \Phi\Phi}^{} -\sigma_{L_iL_i \to \Phi\Phi}^{}
}{\sigma_{\overline{L}_i\overline{L}_i \to \Phi\Phi}^{} +\sigma_{L_iL_i \to \Phi\Phi}^{}}. 
\end{align}
Note that $\epsilon_i^{}$ corresponds to the interaction, where one $L_i$ comes from inflaton decay 
and another one from thermal plasma. 
On the other hand, $\epsilon_{2i}^{}$ corresponds to the collision between leptons both from inflaton decay.
More specifically, $\epsilon$'s are given by
\begin{align}
& \epsilon_{i}^{} \simeq 
\sum_j \frac1{2\pi} \frac{12M_\text{inf}\, T_R}{\Lambda_2^2}
\frac{ \lambda^{(1)}_{jj} \text{Im}(\lambda^{(2)}_{ijij})}{\lambda^{(1)}_{ii}},
& \epsilon_{2i}^{} \simeq 
\sum_j \frac1{8\pi} 
\frac{M_\text{inf}^2}{\Lambda_2^2} 
\frac{\lambda^{(1)}_{jj}\text{Im}(\lambda^{(2)}_{ijij})}{\lambda^{(1)}_{ii}}.
\end{align}
We denote the interaction rates of the lepton number violation process corresponding to $\epsilon_{i}$ and $\epsilon_{2i}$ by $\Gamma_{\cancel{L}_i}$ and $\Gamma_{2\cancel{L}_i}$, respectively:
\begin{align}
& \Gamma_{\cancel{L}_i}\simeq
\frac{11}{4\pi^3}\, \zeta(3)\, {m_{\nu, i}^2\over v^4}\, T^3,
& \Gamma_{2\cancel{L}_i}\simeq
{11\over8\pi}{m_{\nu,i}^2\over v^4}n_{\ell_i}.
\end{align}
The interaction rates of the thermalization process $\Gamma_\text{brems}$, 
and of the washout process $\Gamma_\text{wash}$ are respectively given by 
\begin{align}
&\Gamma_\text{brems}
 \simeq {\alpha_2^2}\, T \sqrt{ \frac{T}{M_\text{inf}}}, \\
&\Gamma_\text{wash}
 \simeq \frac{11}{4\pi^3}\, \zeta(3)\, \frac{\sum m_\nu^2}{v^4}\, T^3,
\end{align}
where $\alpha_2^{}$ is the structure constant of the SU(2)${}_L$ gauge coupling. 

The baryon asymmetry in the reheating era leptogenesis is roughly estimated as~\cite{Hamada:2015xva}, 
\begin{align}
\frac{n_B^{}}{s} 
\simeq&\,  
7.2\times10^{-11} 
\left( \frac{2\times10^{-2}}{\alpha_2} \right)^2 
\left( \frac{T_R}{3\times10^{11}\text{GeV}} \right)^{7/2}
\left( \frac{M_\text{inf}^{}}{2\times10^{13} \text{GeV}} \right)^{1/2} 
\nonumber\\ 
& \qquad 
\times\sum_{i,j}\, \mathcal{B}_i\,\,
\lambda_{ii}^{(1)}\lambda_{jj}^{(1)}
\Big( \frac{6\times10^{14}\text{GeV}}{\Lambda_1} \Big)^2
\text{Im}(\lambda_{ijij}^{(2)})
\Big( \frac{10^{15}\text{GeV}}{\Lambda_2} \Big)^2,
\label{eq:approx}
\end{align}
from which we can see that the observed value of the baryon asymmetry can be reproduced.
%As a result, it is concluded that the observed value of the baryon asymmetry can be reproduced 
%as in Eq.\eqref{eq:approx} in the reheating era leptogenesis scenario. 
%
Let us give a few comments in order.
In the conventional scenario of the leptogenesis, 
the right-handed neutrino on mass-shell decays into leptons in the early universe. 
On the other hand, in the reheating era leptogenesis, 
the right-handed neutrino can be an off-shell particle. 
Thus, it is expected that the allowed region for masses of right-handed neutrinos $M_{R, i}^{}$ and 
the reheating temperature $T_R$ is extended in this new scenario. 
Moreover, the right-handed neutrinos are no longer necessary ingredient of the scenario. 

%============================================================================

\section{The reheating era leptogenesis in models with the seesaw mechanism}
\label{sec3}

\subsection{The Type-I seesaw mechanism}

\begin{figure}[tb]
\centering 
 \includegraphics[height=3.5cm]{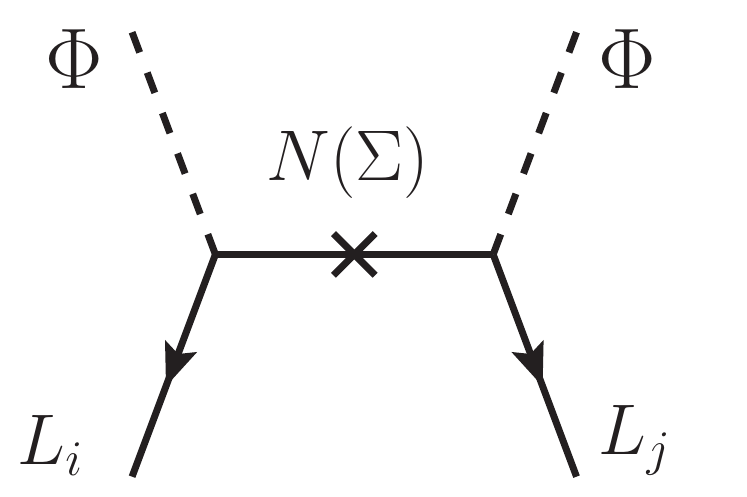} 
  \caption{A process that gives the operator 
           $(\overline{L_i}\widetilde{\Phi})(\overline{L_j}\widetilde{\Phi})/\Lambda_1$ 
           in the Type-I (-\III) seesaw model.} 
\label{fig:dim5op}
\end{figure}

Typical examples of the reheating era leptogenesis are many variations of 
the neutrino mass generation models with the seesaw mechanism. 
A simplest one is the Type-I seesaw model~\cite{typeI}, which is described by the Lagrangian, 
\begin{align}
\Delta{\mathcal L}^\text{Type-I}
= 
+y^\text{I}_{ij}\overline{L_i}{N_R}_j\widetilde{\Phi}
+\frac{M_{R,i}^{}}{2}\overline{{N^c_R}_i}{N_R}_i+\text{H.c.},  
\end{align}
where $N_R$ represents right-handed neutrinos. 
The mass matrix for left-handed neutrinos is generated by the Type-I seesaw mechanism in Fig.~\ref{fig:dim5op},  
which is expressed as 
\begin{align}
m_\nu^{} = -\frac{v^2}2 {y_{}^\text{I}} M_R^{-1}{y_{}^\text{I}}^T. 
\label{eq:type-I}
\end{align} 
Note that the coefficient of the first term in Eq.\eqref{eq:leptogenesis} links to $m_\nu^{}$ by Eq.\eqref{eq:seesaw}, 
and the origin of the lepton number violation is caused by the Majorana mass of the right-handed neutrinos. 

By using the Casas-Ibarra parametrization~\cite{Casas:2001sr}, 
the Yukawa matrix can be written with the active neutrino Majorana masses $m_{\nu, i}^{}$ 
and right-handed neutrino Majorana masses $M_{R,i}^{}$ as 
\begin{align}
y^\text{I}_{ij} = i\, \frac{\sqrt{2}}{v} \sqrt{m_{\nu, i}^{}}\, R_{ij}\, \sqrt{M_{R,j}^{}}, 
\end{align}
where $R$ is a complex orthogonal matrix, which satisfies $RR^T=1$. 
We again note that we work in the real diagonal basis of $m_\nu^{}$ (or equivalently $\lambda^{(1)}$). 
The size of matrix elements of $R$ is arbitrary as long as they are complex parameters, 
but $R_{ij}={\mathcal{O}}(1)$ would be a natural choice if the neutrino mass hierarchy is maintained  
without a fine-tuning in the structure of the Yukawa matrix. 
\begin{figure}[tb]
\centering
    \includegraphics[height=3.5cm]{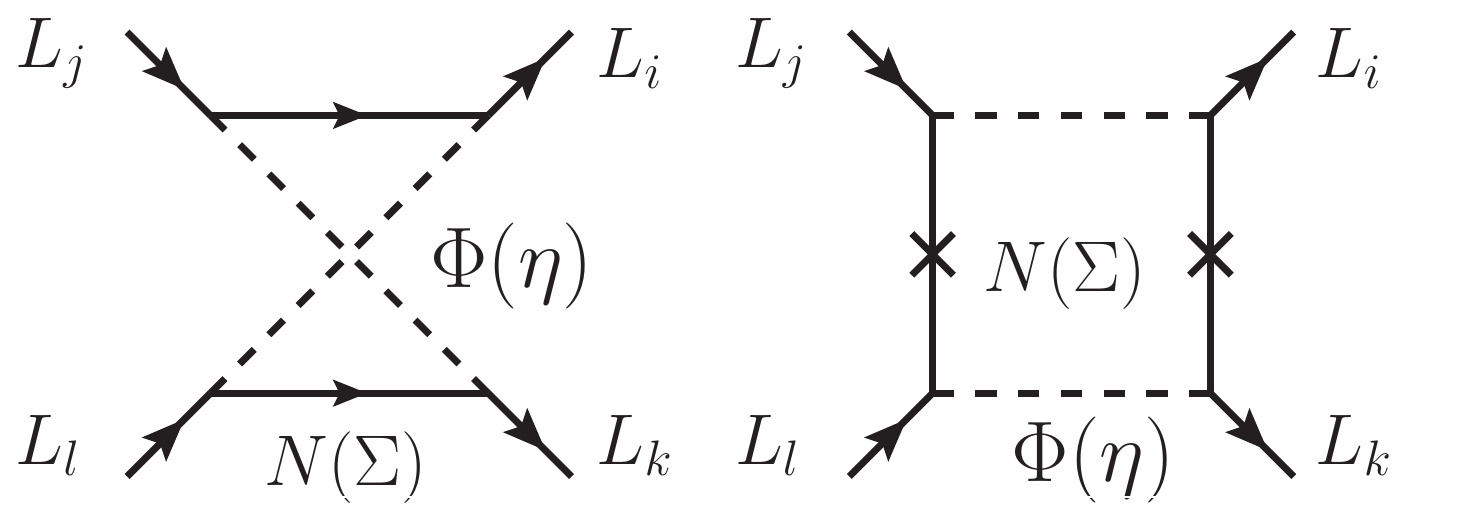}
\caption{Processes that give the operator $(\overline{L_i} \gamma^\mu L_j)(\overline{L_k} \gamma_\mu L_l)/\Lambda_2^2$.} 
\label{fig:dim6op}
\end{figure}
In this framework, the second term in Eq.\eqref{eq:leptogenesis} is also induced 
by the one-loop processes shown in Fig.~\ref{fig:dim6op}. 
The imaginary part of the coefficient of the dimension-six operator can be generated 
only by the left diagram in Fig.~\ref{fig:dim6op}:
\begin{align}
\frac{\text{Im} (\lambda^{(2)}_{ijkl})}{\Lambda_2^2} 
\simeq 
\frac{1}{(8\pi)^2} \sum_{m, n} 
\frac{ \text{Im}(y^\text{I}_{im} y^{\text{I}*}_{lm} y^\text{I}_{kn} y^{\text{I}*}_{jn})}{
M_{R,m}^2-M_{R,n}^2} \log\frac{M_{R,m}^2}{M_{R,n}^2}. 
\label{eq:lambda2}
\end{align}

We are now ready to write down $\lambda^{(1)}_{ij}/\Lambda_1$ and $\text{Im}(\lambda^{(2)}_{ijkl})/\Lambda_2^2$ 
in terms of the parameters in the neutrino sector, i.e., the mass eigenvalues $m_{\nu, i}^{}$ and $M_{R, i}^{}$ and 
a complex orthogonal matrix $R$. 
The baryon asymmetry generated in the reheating era leptogenesis scenario is
roughly evaluated within the framework of the Type-I seesaw model as%~\cite{Hamada:2015xva} 
\begin{align}
\frac{n_B^{}}{s} &=
1.9\times10^{-14}
\left( \frac{2\times10^{-2}}{\alpha_2} \right)^2
\left( \frac{T_R}{10^{11}{\rm GeV}} \right)^{7/2}
\left( \frac{M_{\rm inf}}{2\times10^{13}{\rm GeV}} \right)^{1/2} \nonumber\\ 
             & \qquad \times\sum_{i,j} \mathcal{B}_i
               \left(\frac{m_{\nu, i}^{}}{0.1{\rm eV}}\right)^2
               \left(\frac{m_{\nu, j}^{}}{0.1{\rm eV}}\right)^2{\rm Im}[(RR^\dag)_{ij}^2]. %, 
\label{eq:nB}
\end{align}
Here and hereafter, we take the degenerate mass limit of right-handed neutrinos, 
$M_{R, 1}^{}=M_{R, 2}^{}=M_{R, 3}^{}$ for simplicity.\footnote{
Even when we consider mass differences among right-handed neutrinos, 
the result of the calculation in this section does not change much. 
In the case with mass differences, sub-leading contributions to $\text{Im}[(RR^\dag)^2_{ij}]$ 
are received a logarithmic correction factor, $\log(M_{R,m}^2/M_{R,n}^2)$}.
In the numerical analysis, the neutrino mass squared differences are chosen as  
$\Delta m_{\nu 21}^2 \equiv m_{\nu, 2}^2-m_{\nu, 1}^2 =7.53\,(7.53)\times10^{-5} \text{eV}^2$ and 
$\Delta m_{\nu 32}^2 \equiv |m_{\nu, 3}^2-m_{\nu, 2}^2| =2.44\,(2.52)\times10^{-3} \text{eV}^2$ 
for the normal (inverted) mass ordering~\cite{Gando:2013nba}. 

For the justification of the effective Lagrangian description in Eq.\eqref{eq:leptogenesis} in our analysis, 
$M_R$ must be heavy enough not to be generated at the on-shell in the early universe. 
This requirement leads to a condition, 
\begin{align}
M_{\rm inf} \hspace{0.3em} \raisebox{0.4ex}{$<$}\hspace{-0.75em}\raisebox{-.7ex}{$\sim$}\hspace{0.3em} M_R. 
\label{eq:lowerbound}
\end{align}
We note that, in Ref.~\cite{Hamada:2015xva}, the upper bound on $M_{\rm inf}$ is not imposed because the ultraviolet completion is not specified.
In order to estimate the lower bound on $T_R$, we choose $M_{\rm inf}$ so as to maximize the baryon asymmetry. 
It can be seen that the asymmetry increases for the larger value of $M_{\rm inf}$ in Eq.~\eqref{eq:nB}. 
Thus, Eq.~\eqref{eq:lowerbound} is regarded as the upper bound on $M_{\rm inf}$. 
Requiring that the gravity does not become strong, we impose another upper bound as $M_\text{inf}\lesssim M_\text{Pl}$. 
Since a large value of $M_\text{inf}$ leads $\epsilon_i^{} (\epsilon_{2i}^{}) \gtrsim 1$, 
we demand the consistency conditions on $M_\text{inf}\lesssim M_1 (M_2)$, where 
$M_\text{inf} = M_1 (M_2)$ is the solutions of $\epsilon_i^{} (\epsilon_{2i}^{})=1$. 
Therefore, we put $M_{\rm inf}=\text{Min}\left( M_R, M_\text{Pl}, M_1, M_2 \right)$ in the following discussions, 
and evaluate the lower bound on $T_R$ for various $M_R$.

\begin{figure}[tb]
\centering
 \includegraphics[width=7.1cm]{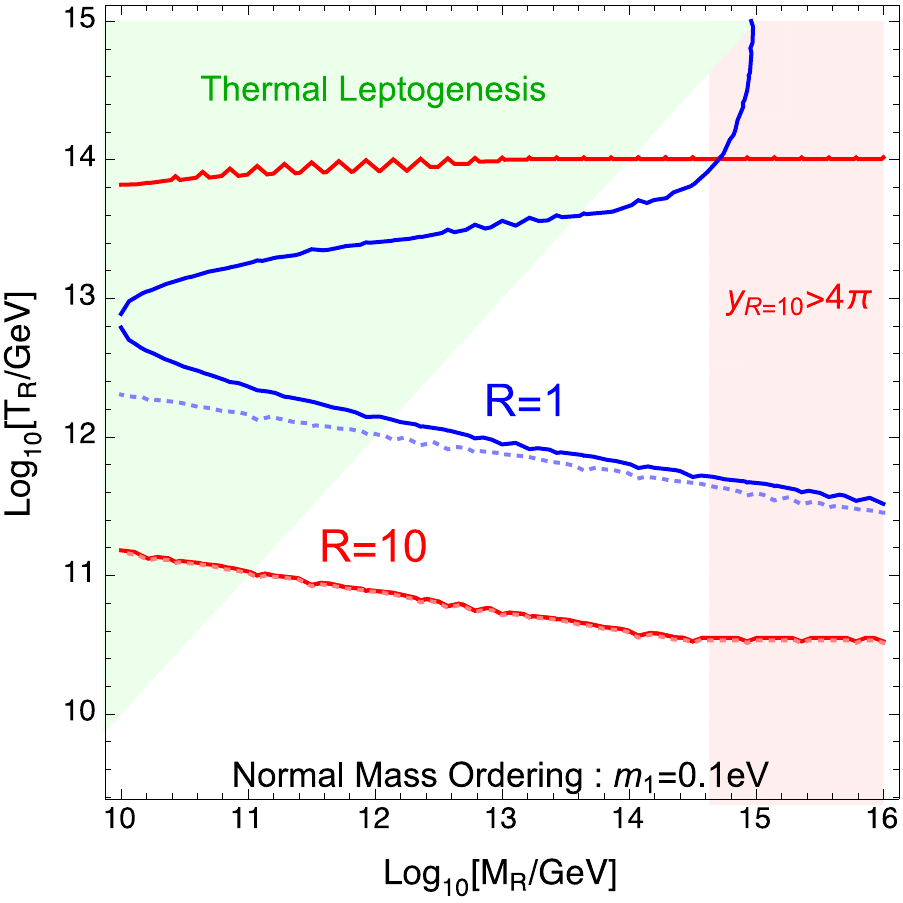} 
 \includegraphics[width=7.1cm]{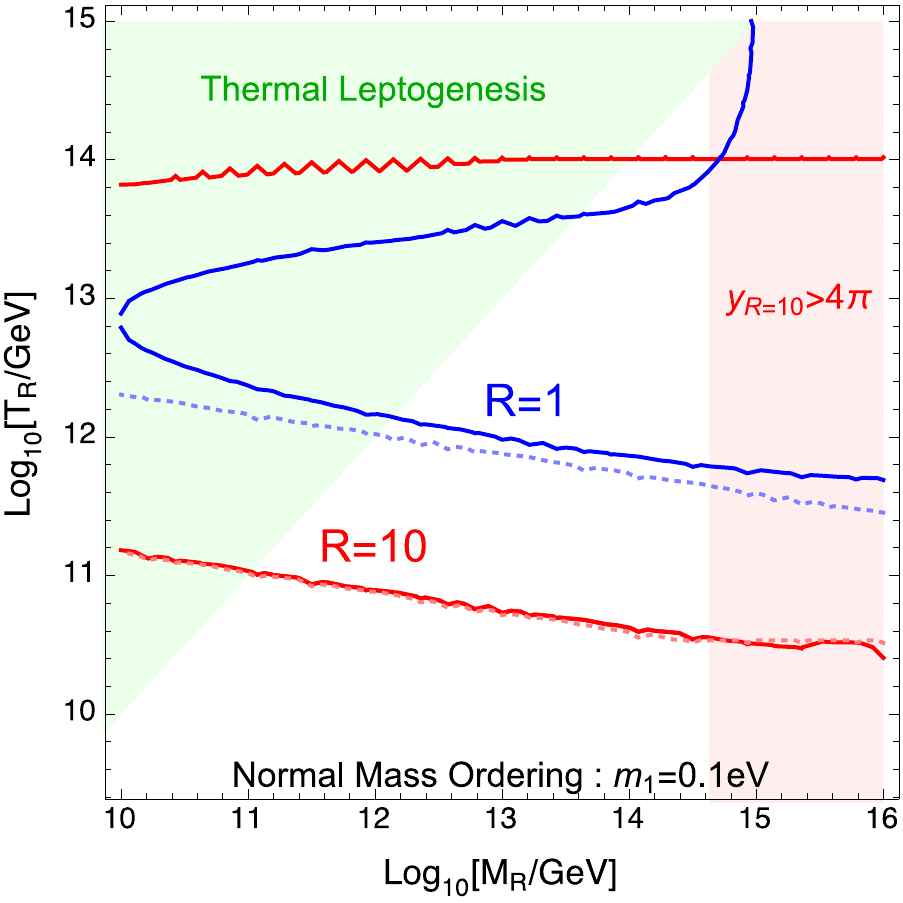}
  \caption{
 The allowed parameter space of $T_R$ as a function of $M_R$ in the Type-I seesaw model for $R=1$ and $R=10$ with normal mass ordering. In the right panel, the effect of $\epsilon_2 $ term is omitted in Boltzmann equation. } 
\label{fig:type-1-NH-highmass}
\end{figure}
%
%\begin{figure}[tb]
%\centering
% \includegraphics[width=7.1cm]{Type1-NH-HighMass.eps} 
% \includegraphics[width=7.1cm]{Type1-IH-HighMass.eps}
%  \caption{
%The allowed parameter space of $T_R$ as a function of $M_R$ in the Type-I seesaw model.
%  } 
%\label{fig:type-1-highmass}
%\end{figure}
%
%
%\begin{figure}[tb]
%\centering
% \includegraphics[width=7.1cm]{Type1-NH-HighMass.eps} 
% \includegraphics[width=7.1cm]{Type1-IH-HighMass.eps}
%  \caption{
%  Lower bound on $T_R$ as a function of $M_R$ in the Type-I seesaw model.
%  } 
%%\label{fig:type-1-highmass}
%\end{figure}
%
In the left panel of Fig.~\ref{fig:type-1-NH-highmass}, the two-dimensional lower bounds are shown in the $T_R$ and the $M_R$ plane for the reheating era leptogenesis. 
In order to see the effect of the newly added $\epsilon_2$ term compared with the letter article~\cite{Hamada:2015xva}, 
we show the lower and upper bounds on $T_R$ without $\epsilon_2$ term in the right panel of Fig.~\ref{fig:type-1-NH-highmass}.
We confirm that the effect of $\epsilon_2$ slightly enlarge the allowed parameter space. 
More concretely, in the left panel, the lower bound on $T_R$ is slightly smaller than that of right panel.
In both cases, we set $m_{\nu1}^{}=0.1$ eV, and $\mathcal{B}_i \propto |U_{\tau i}|^2, \sum_i\mathcal{B}_i=1$ as in footnote 3.
Then, we have
\begin{align}
&
\mathcal{B}_1\simeq0.19,
&&
\mathcal{B}_2\simeq0.25,
&&
\mathcal{B}_3\simeq0.56.
\label{eq:branching}
\end{align}
Here we take the observed values of mixing angles, a maximum Dirac phase~\cite{Abe:2013hdq}, and vanishing Majorana phases.
The solid-blue (-red) curve expresses the numerical results with the magnitude of the matrix elements to be $R_{ij} = 1\, (10)$. 
To be precise, the following relations are adopted, 
$R^2 \equiv {\rm Im}[(RR^\dag)_{12}^2]={\rm Im}[(RR^\dag)_{13}^2]={\rm Im}[(RR^\dag)_{23}^2]
=-{\rm Im}[(RR^\dag)_{12}^2]=-{\rm Im}[(RR^\dag)_{31}^2]=-{\rm Im}[(RR^\dag)_{32}^2]$. 
Upper-right regions of the curves are allowed parameter space for the successful leptogenesis. 
Note that the contributions from the decay of the right-handed neutrinos are not included in our analysis, 
instead, we indicate the corresponding parameter space $T_R \gtrsim M_R$ (upper-left domain), 
where the thermal leptogenesis would be realized. 
The shaded region in larger $M_R$ indicates the breakdown of the perturbativity for the Yukawa coupling, 
which is defined by $y_{\nu}^{}(R=10)>4\pi$. For $R=1$, the perturbativity condition is satisfied in all 
the parameter regime in the plot. For $R=10$, there exists the upper bound on $T_R$ because of the strong washout.
We notice that the condition $\epsilon_2\lesssim1$ is numerically almost close to the perturbativity condition of the Yukawa coupling.

The dotted lines represent the analytic result in Eq.\eqref{eq:nB}. 
Combining with Eq.\eqref{eq:nB} and $M_{\rm inf}=M_R$, 
the behavior of the lower bound on the reheating temperature is $T_R \propto M_R^{-1/7}$.  
For larger $M_R$, the lower bound on $T_R$ is approximately constant, 
since we convolute Eq.~\eqref{eq:nB} and $M_{\rm inf}=M_2$. 
You can see our numerical results are well consistent with the approximated results including the overall factor. 
For very large $T_R$ and relatively small $M_R$ region, the effect of the washout becomes important  
so that a corner of the parameter space is not suitable for the leptogenesis. 
For both $T_R$ and $M_R$ large region, because $M_\text{inf}$ is strongly constrained by the condition $\epsilon_i <1$, 
the maximally produced baryon asymmetry is not enough for explaining our universe. 

\begin{figure}[tb]
\centering
 \includegraphics[width=7.1cm]{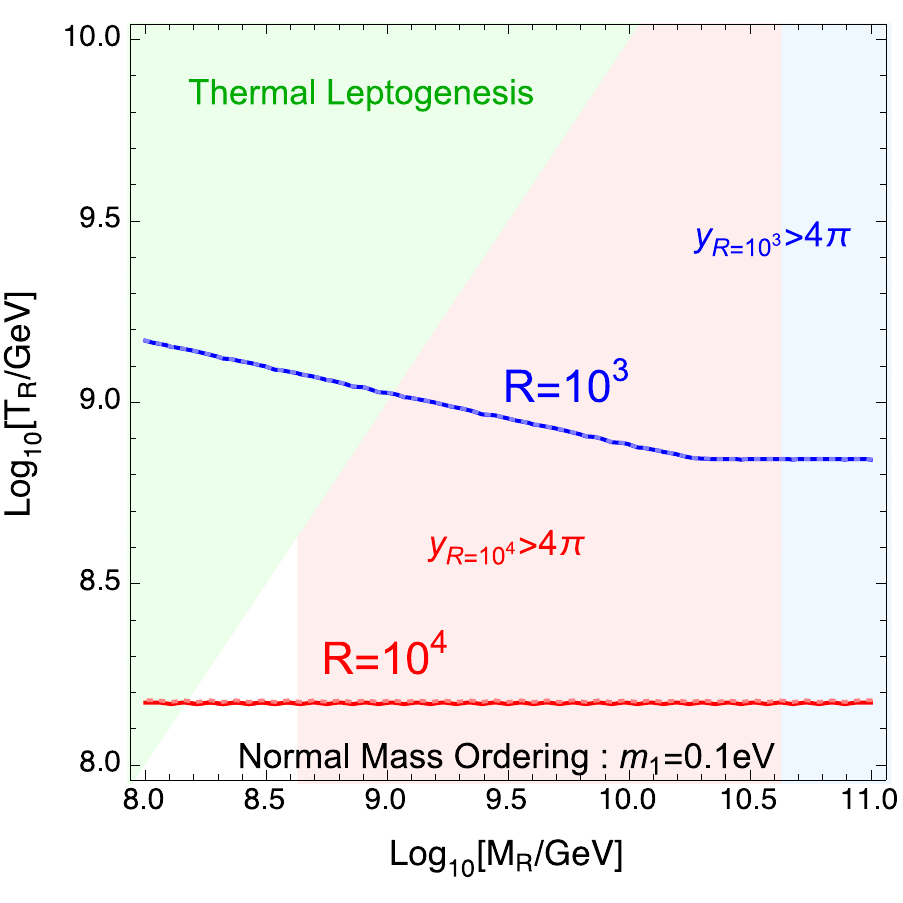} 
 \includegraphics[width=7.1cm]{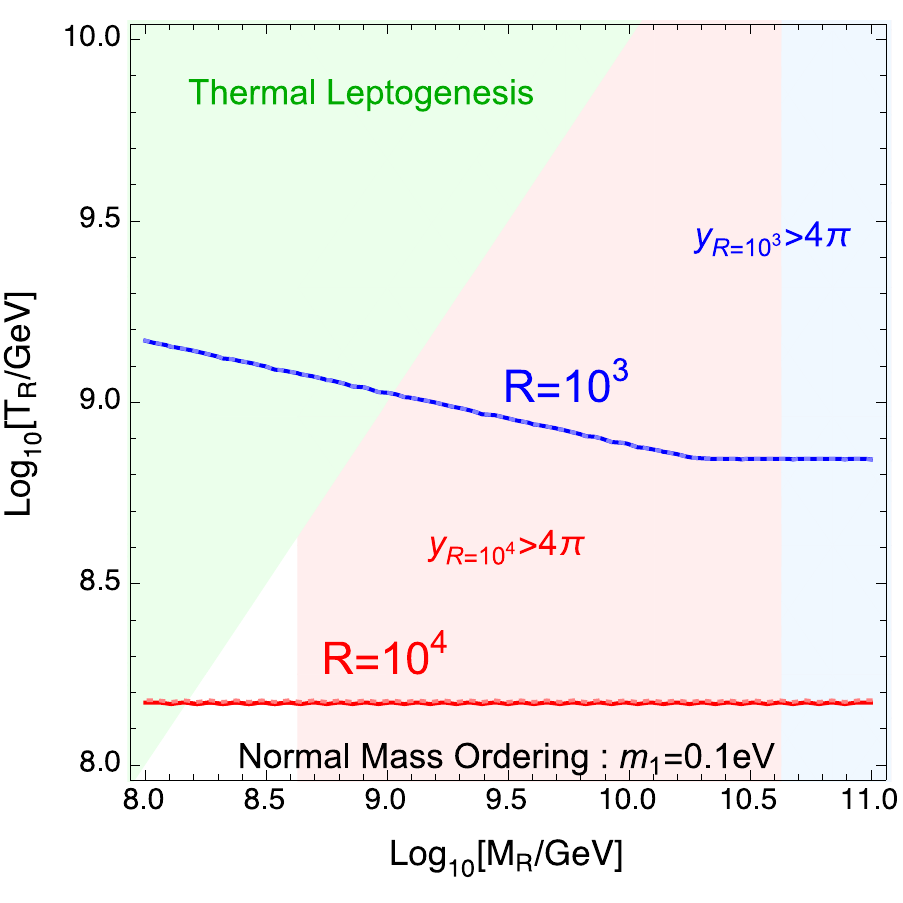}
  \caption{
 The allowed parameter space of $T_R$ as a function of $M_R$ in the Type-I seesaw model for $R=10^3$ and $R=10^4$ with normal mass ordering. In the right panel, the effect of $\epsilon_2 $ term is omitted in Boltzmann equation.  } 
\label{fig:type-1-NH-lowmass}
\end{figure}

\begin{figure}[tb]
\centering
 \includegraphics[width=7.1cm]{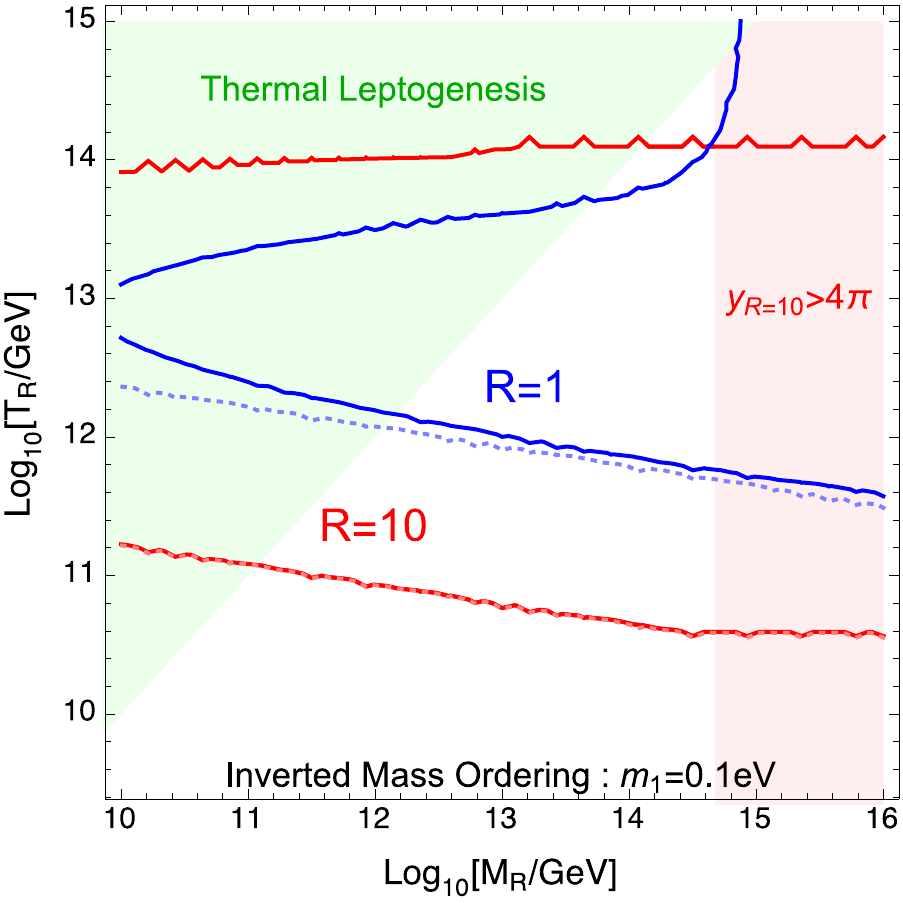} 
 \includegraphics[width=7.1cm]{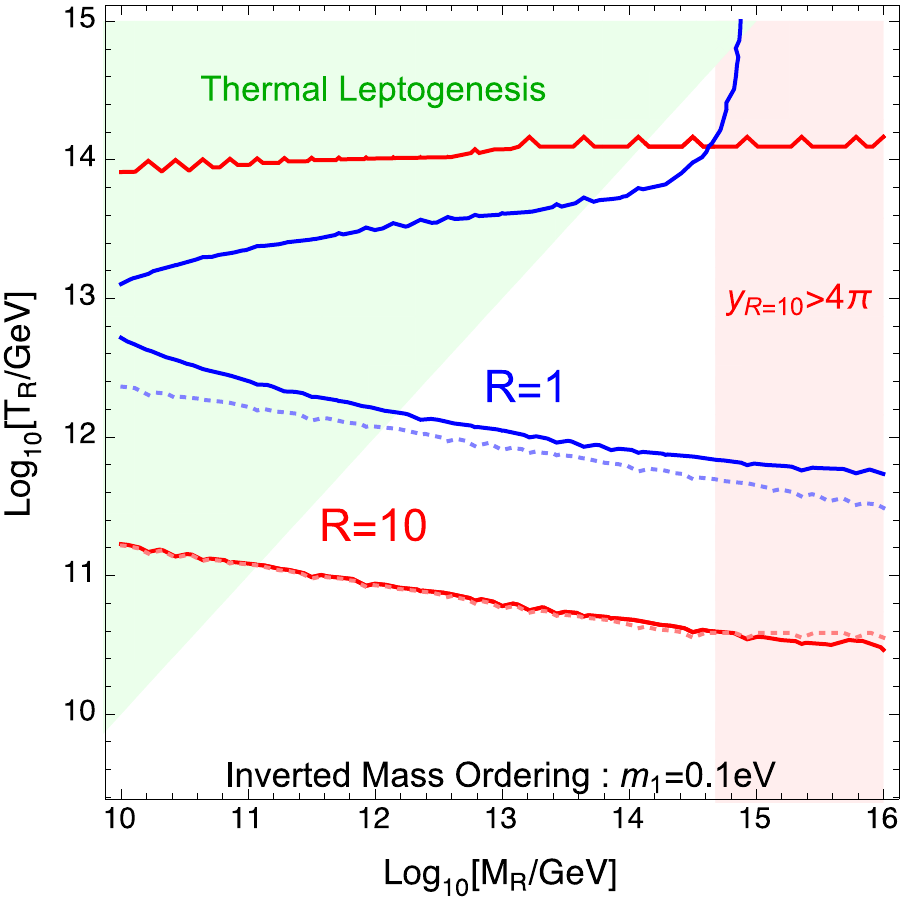}
  \caption{
 The allowed parameter space of $T_R$ as a function of $M_R$ in the Type-I seesaw model for $R=1$ and $R=10$ with inverted mass ordering. In the right panel, the effect of $\epsilon_2 $ term is omitted in Boltzmann equation.}  
\label{fig:type-1-IH-highmass}
\end{figure}

\begin{figure}[tb]
\centering
 \includegraphics[width=7.1cm]{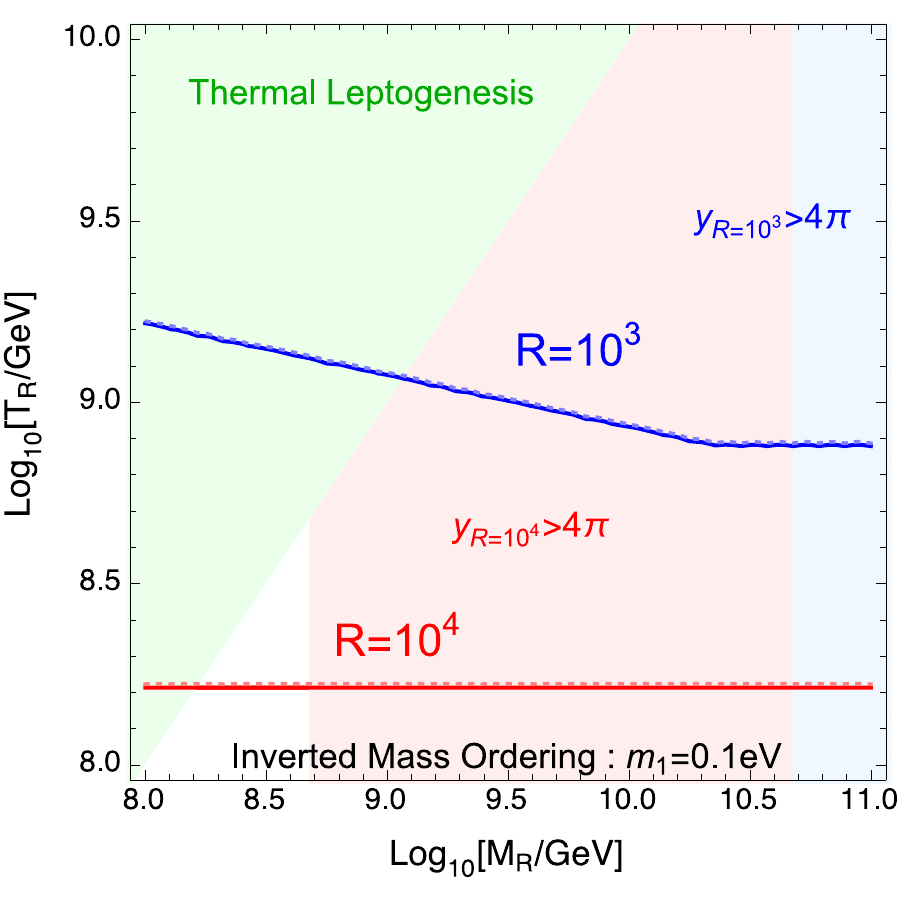} 
 \includegraphics[width=7.1cm]{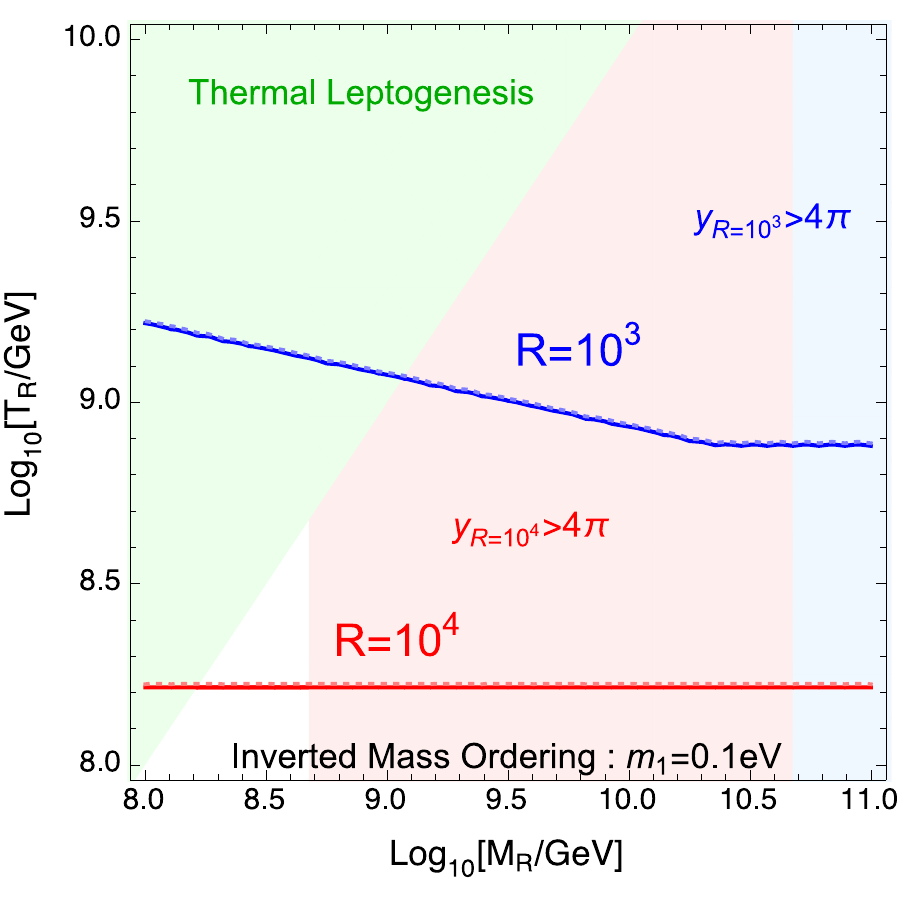}
  \caption{
 The allowed parameter space of $T_R$ as a function of $M_R$ in the Type-I seesaw model for $R=1$ and $R=10$ with inverted mass ordering. In the right panel, the effect of $\epsilon_2 $ term is omitted in Boltzmann equation.  } 
\label{fig:type-1-IH-lowmass}
\end{figure}

In Fig.~\ref{fig:type-1-NH-lowmass}, we show the similar plots but for $R=10^3$ and $R=10^4$. 
The results for the inverted mass ordering of active neutrino masses are shown in Figs.~\ref{fig:type-1-IH-highmass} and \ref{fig:type-1-IH-lowmass}.
The lines and shaded regions are given in the same manner as in Fig.~\ref{fig:type-1-NH-highmass}.\footnote{
For the hierarchical right handed neutrino mass, thermal leptogenesis works only for $T_R\gtrsim10^{10}$GeV and $M_R\gtrsim10^{9}$GeV~\cite{Davidson:2002qv}.
Below these values, the degeneracy of the mass of the right handed neutrino is required~\cite{Pilaftsis:2003gt}.
} 
It is allowed parameter space, but it might be necessary to introduce a fine-tuning among the parameters. 
If we take such a large $R$, the reheating temperature decreases up to about $10^8$ GeV. 
This result will be compared with the case in the radiative seesaw model, where 
the fine-tuning issue can be replaced by a natural small parameter.
For $R=10^5$, all the parameter space is excluded by the perturbativity constraint. 
%
%\begin{figure}[tb]
%\centering
% \includegraphics[width=7.1cm]{Type1-NH-LowMass.eps} 
% \includegraphics[width=7.1cm]{Type1-IH-LowMass.eps}
%  \caption{
%  The allowed parameter space of $T_R$ as a function of $M_R$ in the Type-I seesaw model for $R=10^3$ and $R=10^4$. 
%  } 
%\label{fig:type-1-lowmass}
%\end{figure}

%\begin{figure}[tb]
%\centering
% \includegraphics[width=7.1cm]{Type1-NH-HighMass-woe2.eps} 
% \includegraphics[width=7.1cm]{Type1-NH-LowMass-woe2.eps}
%\caption{
%The allowed parameter space of $T_R$ as a function of $M_R$ in the Type-I seesaw model. The effect of $\epsilon_2 $ term is omitted in Boltzmann equation.
%} 
%\label{fig:type-1 woe2}
%\end{figure}
%
%
%\begin{figure}[tb]
%\centering
% \includegraphics[width=7.1cm]{Type1-NH-LowMass-kai.eps} 
% \includegraphics[width=7.1cm]{Type1-IH-LowMass-kai.eps}
%  \caption{
%  Lower bound on $T_R$ as a function of $M_R$ in the Type-I seesaw model for $R=10^3$ and $R=10^4$. 
%  } 
%%\label{fig:type-1-lowmass}
%\end{figure}
%

\subsection{The Type-\III\,  seesaw mechanism}

The Type-I\hspace{-.1em}I\hspace{-.1em}I seesaw model is one of variations of the tree-level seesaw mechanism.\footnote{There is one more tree-level seesaw mechanism. 
In the Type-II, an SU(2)${}_L$ triplet scalar $\Delta$ is introduced. 
The new Yukawa interaction $\overline{L^c}i\,\sigma_2 \Delta L$ is the origin of Majorana neutrino masses 
when $\Delta$ develops VEV. Since the new Yukawa matrix is simultaneously diagonalized 
with the neutrino mass matrix, no new CP violating phase is provided. Thus, the leptogenesis does not work 
in this minimal setup. }
Instead of the SU(2)${}_L$ singlet right-handed neutrinos in the Type-I seesaw model, 
the SU(2)${}_L$ triplet fields $\Sigma$ are added to the SM. The Lagrangian is described as 
\begin{align}
\Delta\mathcal{L}^\text{Type-\III} = 
+y^\text{\III}_{ij} \overline{(L_i)_\alpha}\sigma^a_{\alpha\beta}\Sigma^a_j(\widetilde{\Phi})_\beta 
+\frac{M_{R,i}}{2}\Sigma^{aT}_iC\Sigma^a_i 
+\text{H.c.}
\end{align}
From this Lagrangian, the left-handed neutrino masses are generated by the Type-III seesaw mechanism as 
\begin{align}
m_{\nu, i}^{} = -\frac{v^2}2\, {y_{}^\text{\III}} M_R^{-1}{y_{}^\text{\III}}^T,
\label{eq:type-III}
\end{align} 
while the imaginary part of the coefficient of the dimension-six operator is given by
\begin{align}
\frac{\text{Im} (\lambda^{(2)}_{ijkl})}{\Lambda_2^2} 
\simeq 
\frac{1}{(8\pi)^2} \sum_{m, n} 
\frac{ \text{Im}(y^\text{\III}_{in} y^{\text{\III}*}_{ln} y^\text{\III}_{km} y^{\text{\III}*}_{jm} 
-4y^\text{\III}_{in} y^{\text{\III}*}_{jn} y^\text{\III}_{km} y^{\text{\III}*}_{lm})}{
M_{R, m}^2-M_{R, n}^2} \log\frac{M_{R,m}^2}{M_{R,n}^2}.  
\label{eq:lambda2-III}
\end{align}
Taking the element $\lambda^{(2)}_{ijij}$, we see that a factor of $3$ enhancement is found  
for the baryon number asymmetry as compared to the Type-I seesaw model with the same parameter choices.

\subsection{The scotogenic seesaw mechanism}

As an example of the different types of the seesaw mechanism, 
we here consider a simple radiative seesaw model proposed in Ref.\cite{ref:ma}. 
An advantage of the radiative seesaw mechanism is that the smallness of neutrino masses 
can be understood not only by heavy particles but also loop suppression factors. 
On the other hand, at least two more new particles are required. 
The Lagrangian for the neutrino mass generation sector in the scotogenic model\cite{ref:ma} is given by
\begin{align}
\Delta{\mathcal L} = 
 y^\text{D}_{ij}\overline{L_i}{N_R}_j\widetilde{\eta} 
+\frac{M_{R,i}}{2}\overline{{N^c_R}_i}{N_R}_i 
+\frac{\lambda_5}{2}(\eta^\dag \Phi)^2 
+\text{H.c.}, 
\end{align}
where a scalar doublet $\eta$ is added to the Type-I seesaw model. 
In addition, an ad-hoc $Z_2$ parity is assumed under which only $N_{R, i}$ and $\eta$ are transformed as odd. 
This discrete symmetry forbids the VEV of $\eta$, and therefore the tree-level neutrino masses are forbidden. 
\begin{figure}[tbp]
\centering
 \includegraphics[height=4cm]{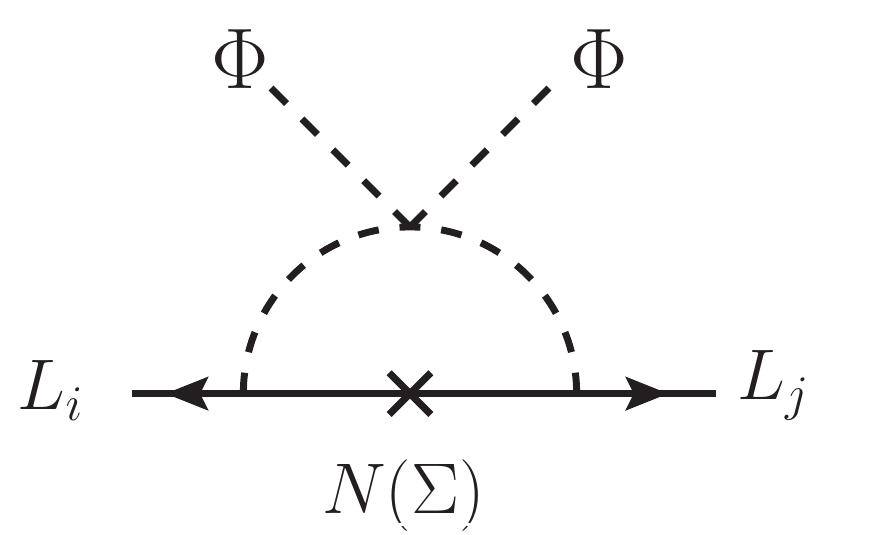} 
  \caption{In the scotogenic radiative seesaw mechanism, 
           $(\overline{L_i}\widetilde{\Phi})(\overline{L_j}\widetilde{\Phi})/\Lambda_1$
           is derived by loop processes.} 
\label{fig:dim5op-2}
\end{figure}
From the one-loop diagram in Fig.~\ref{fig:dim5op-2}, masses of left-handed neutrinos are generated as 
\begin{align}
m_{\nu, i}^{} 
%= 
%-\frac{v^2}{2}\frac{\lambda_5}{(2\pi)^2} 
%y^\text{D} \frac{M_R}{M_R^2-M_\eta^2} 
%\left(1-\frac{M_R^2}{M_R^2-M_\eta^2} \log \frac{M_R^2}{M_\eta^2}\right)
%{y^{\text{D}}}^T 
\equiv 
-\frac{v^2}{2}y^\text{D} {M_R^\text{eff}}^{-1} {y^{\text{D}}}^T, 
\end{align}
where the effective right-handed neutrino mass matrix $M_R^\text{eff}$ is defined as 
\begin{align}
{M_R^\text{eff}}^{-1} = 
\frac{\lambda_5}{(2\pi)^2}\, F({M_R^2}/M_\eta^2)\, M_R^{-1}, 
\quad F(x)=\frac{x}{x-1} \Big(\frac{x}{x-1}\log{x}-1\Big). 
\label{eq:mReff2}
\end{align}
The mass of $\eta$ is $M_\eta$, and the parameter $\lambda_5$ 
characterizes the mixing between the CP even and odd neutral components of $\eta$. 
The coefficient $\lambda^{(2)}_{ijkl}$ of the dimension-six operator in the scotogenic model is 
calculated similarly to as that in the Type-I seesaw model, 
where the Higgs doublet in Fig.~\ref{fig:dim6op} is simply replaced by $\eta$. 
As long as $M_\eta \ll M_R$, $\lambda^{(2)}_{ijkl}$ is the same as Eq.\eqref{eq:lambda2}
substituting $y^\text{I}_{}$ by $y^\text{D}_{}$. 
 
Similarly to the Type-I seesaw mechanism, the Yukawa matrix $y_{}^\text{D}$ is expressed as 
\begin{align}
y^\text{D}_{ij} = i\, \frac{\sqrt{2}}{v} \sqrt{m_{\nu, i}^{}}\, R_{ij}\, \sqrt{M_{R,j}^\text{eff}}. 
\end{align}
Note that the magnitude of the Yukawa coupling can be much larger than that in the Type-I while keeping $R={\mathcal O}(1)$, 
because an additional loop suppression factor $(2\pi)^2$ and a possible small coupling $\lambda_5$ 
are contained in $M_{R}^\text{eff}$. 
In fact, the smallness of $\lambda_5$ can be justified by the naturalness argument, 
since $\lambda_5$ is a lepton number violating parameter if we assign the lepton number of $\eta$ to be unity 
instead of the right-handed neutrinos. 
Namely, the lepton number symmetry is recovered in the $\lambda_5\to 0$ limit. 
For the model building, see Ref.~\cite{Ref:HTT} for example. 

The lower bound on $T_R$ in the scotogenic model is easily estimated the corresponding analytic formula. 
Comparing the result in the Type-I seesaw, we find that 
\begin{align}
T_R^\text{scotogenic}\, \Big( \frac{\lambda_5}{(2\pi)^2}\, F(M_R^2/M_\eta^2) \Big)^{-4/7}
\simeq T_R^\text{Type-I},  
\label{eq:TR}
\end{align}
for smaller $M_R$, where we set $R=1$ both in the scotogenic and the Type-I seesaw models. 
This simple relation suggests that the fine-tuning of $R$ in the Type-I seesaw can be replaced 
by the smallness of $\lambda_5$ in the scotogenic model. 

%\begin{figure}[tb]
%\centering
% \includegraphics[width=7.1cm]{Ma-NH-HighMass.eps} 
% \includegraphics[width=7.1cm]{Ma-IH-HighMass.eps}
% \includegraphics[width=7.1cm]{Ma-NH-LowMass.eps} 
% \includegraphics[width=7.1cm]{Ma-IH-LowMass.eps}
%\caption{Lower bound on $T_R$ as a function of $M_R$ in the Ma's radiative seesaw model.} 
%\label{fig:ma}
%\end{figure}

\begin{figure}[tb]
\centering
 \includegraphics[width=7.1cm]{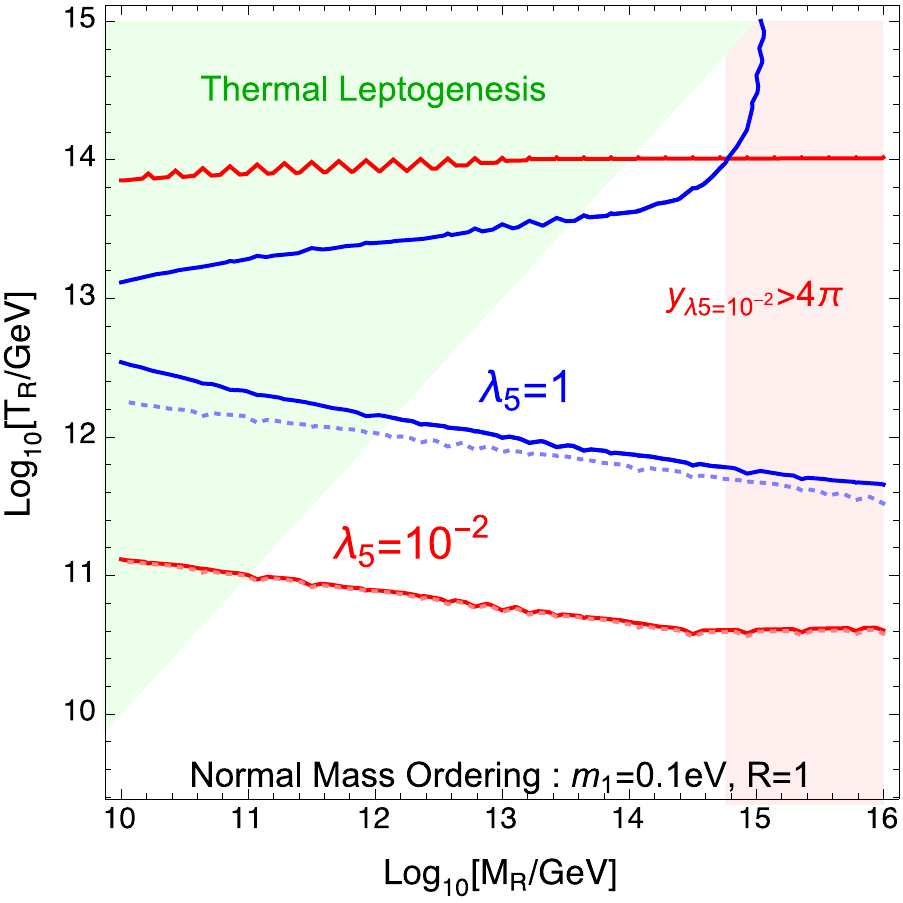} 
 \includegraphics[width=7.1cm]{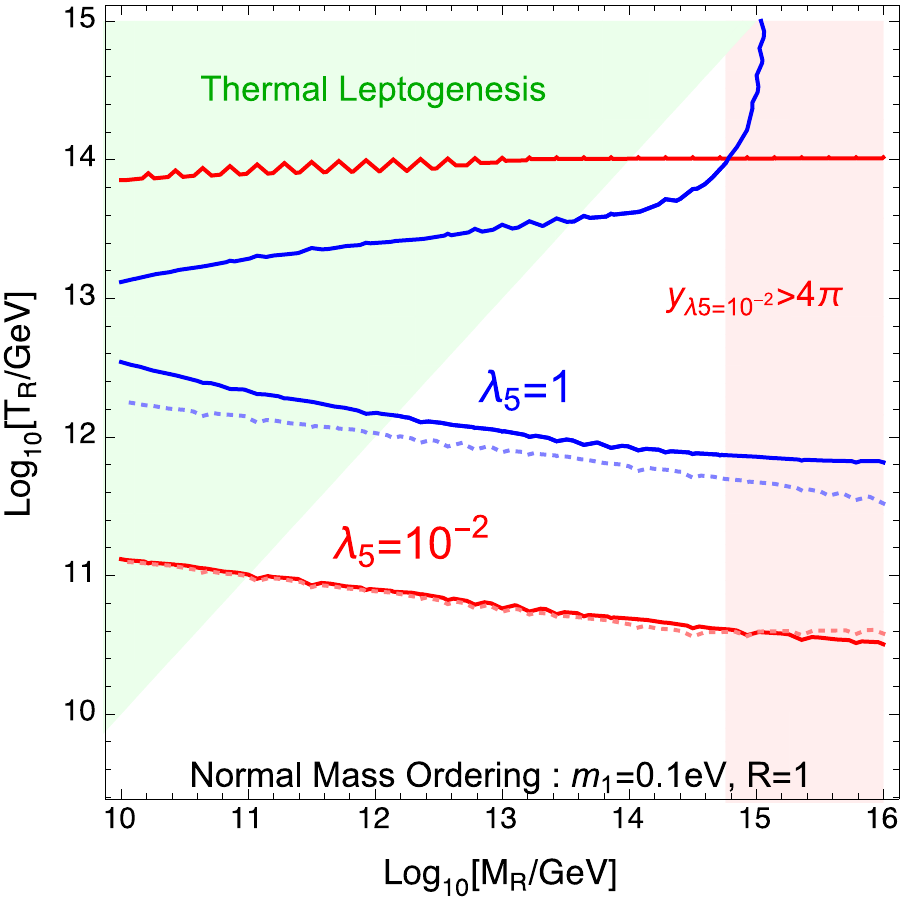}
 \includegraphics[width=7.1cm]{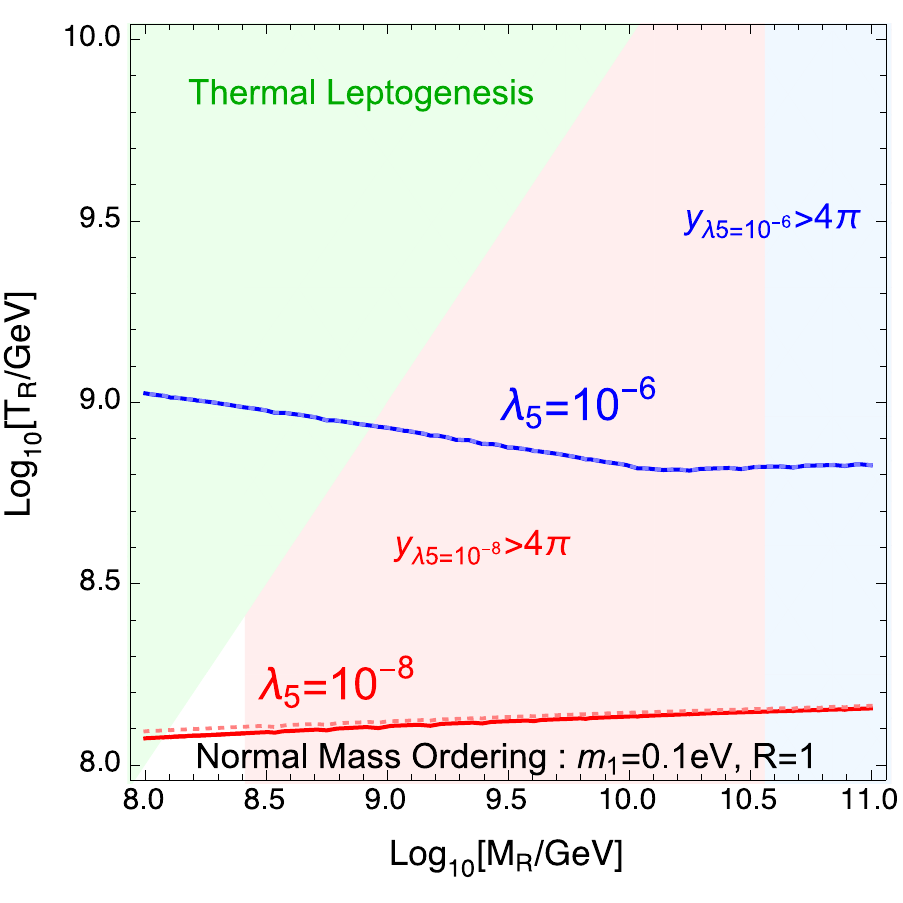} 
 \includegraphics[width=7.1cm]{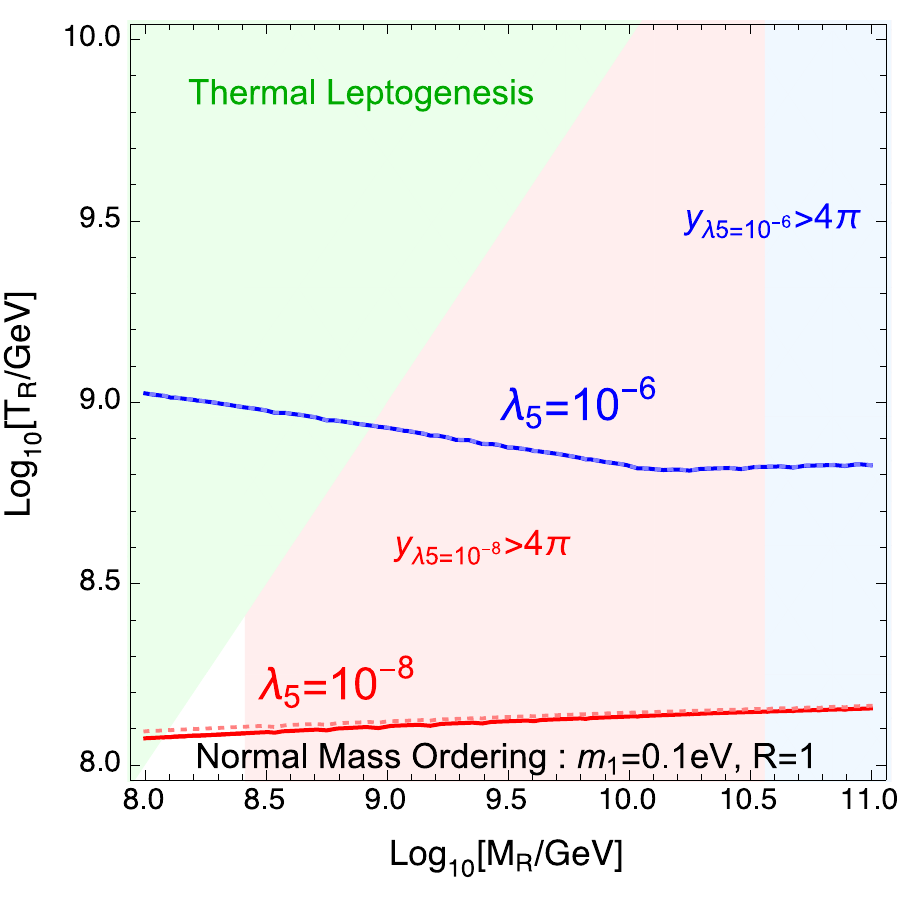}
\caption{ The allowed parameter space of $T_R$ as a function of $M_R$ in Ma model with normal mass ordering. In the right panel, the effect of $\epsilon_2 $ term is omitted in Boltzmann equation.} 
\label{fig:ma-NH}
\end{figure}

\begin{figure}[tb]
\centering
 \includegraphics[width=7.1cm]{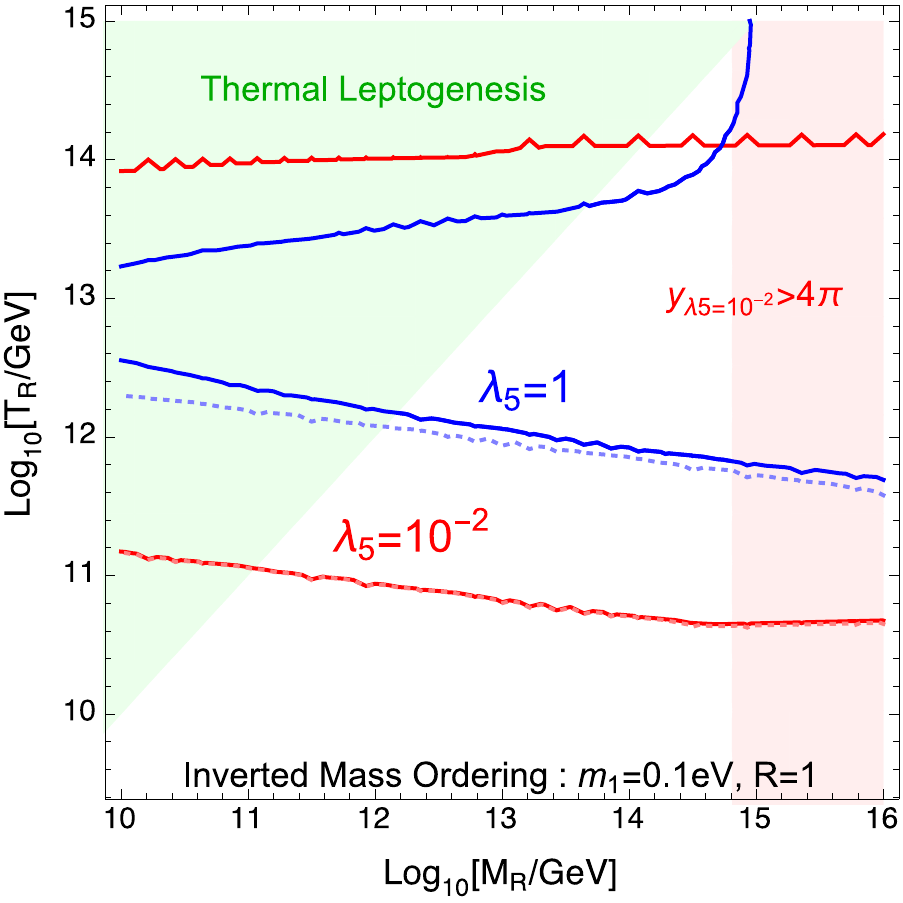} 
 \includegraphics[width=7.1cm]{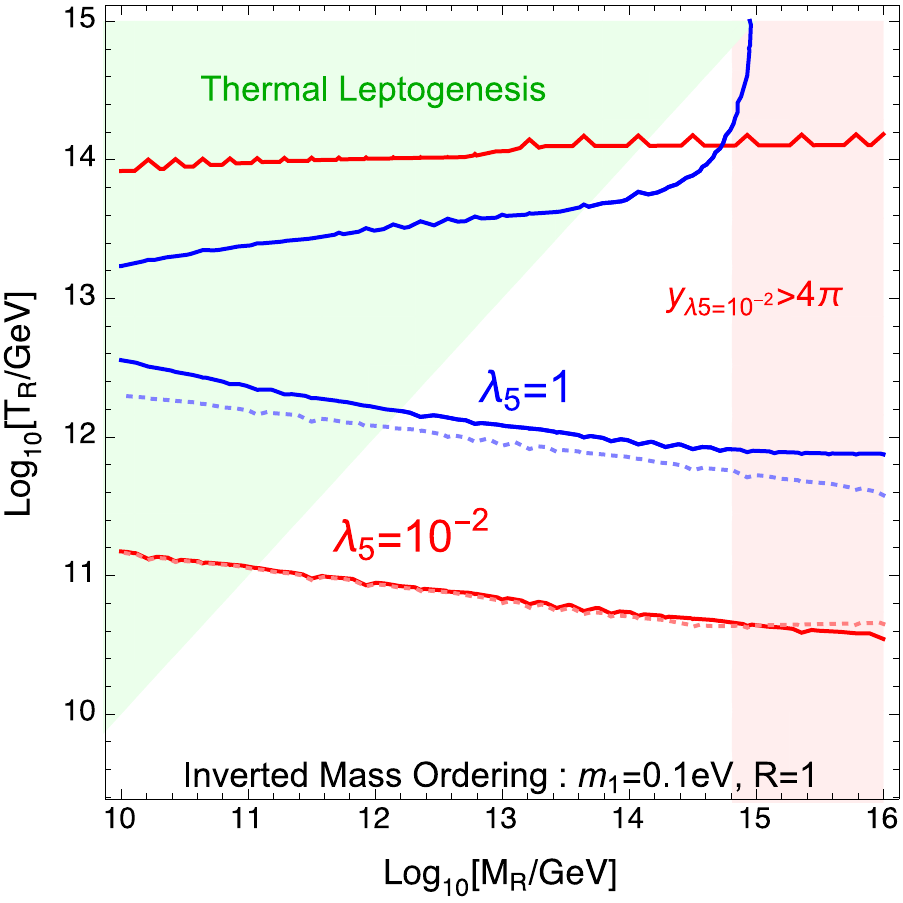}
 \includegraphics[width=7.1cm]{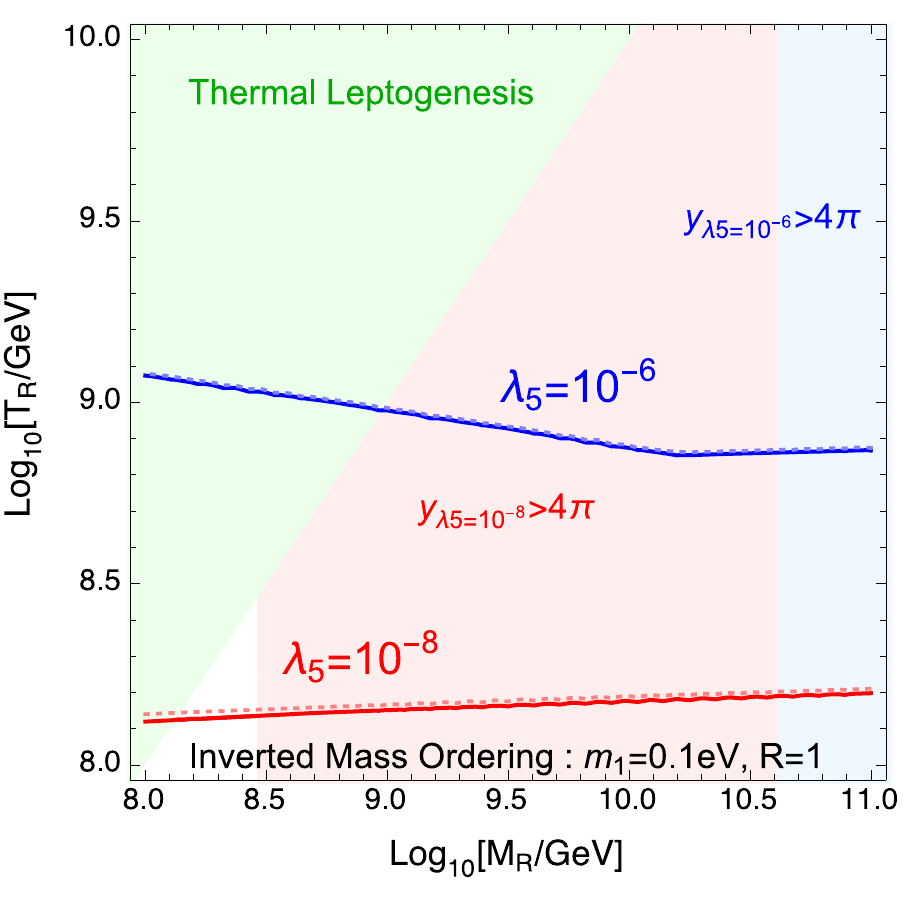} 
 \includegraphics[width=7.1cm]{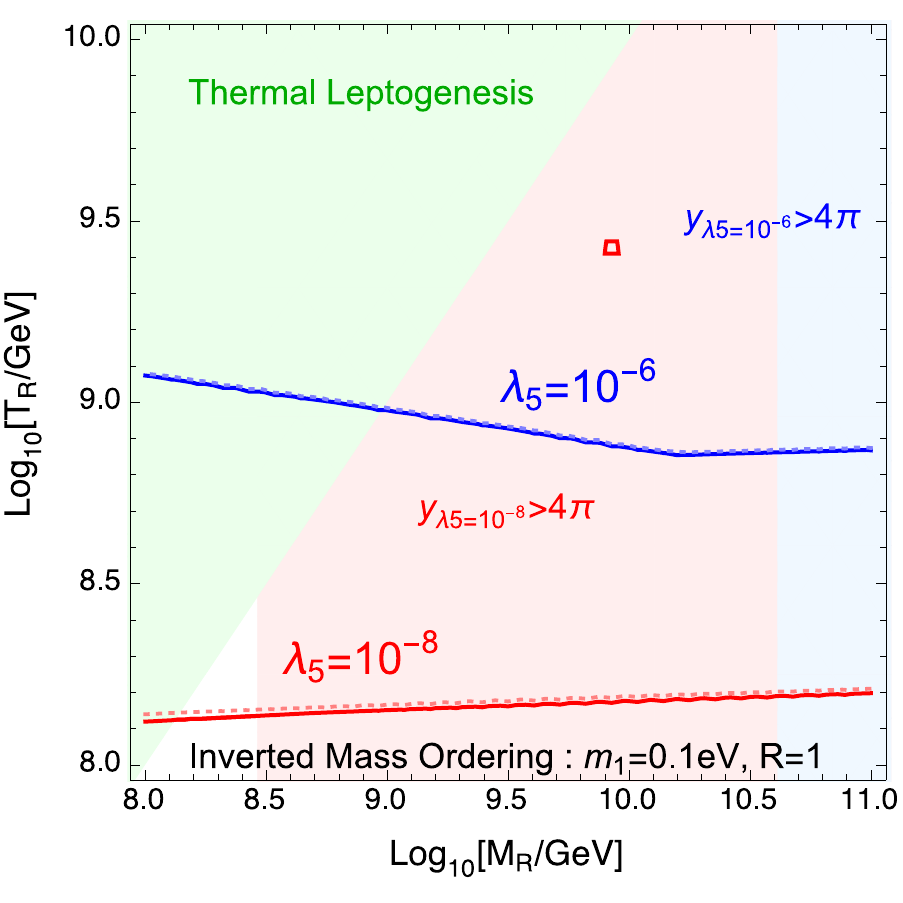}
\caption{ The allowed parameter space of $T_R$ as a function of $M_R$ in Ma model with normal mass ordering. In the right panel, the effect of $\epsilon_2 $ term is omitted in Boltzmann equation.} 
\label{fig:ma-IH}
\end{figure}

Now, we are ready for examining how the allowed region for $T_R$ is extended 
in the Ma's radiative seesaw model. 
In the left panels of Figs.~\ref{fig:ma-NH} and \ref{fig:ma-IH}, the lower bounds on $T_R$ as a function of $M_R$ are shown. 
Fig.~\ref{fig:ma-NH} (\ref{fig:ma-IH}), we show the results for the normal (inverted) mass ordering of active neutrino masses. 
The curves and shaded regions are given in the similar manner as the plots of the Type-I seesaw model.  
The mass of the inert doublet is chosen to be $M_\eta=10^3$ GeV, 
which is not sensitive to the numerical analysis if $M_\eta \ll M_R$. 
The magnitude of $R$ matrix elements is $R=1$ in all the plot. 
Instead, we take different values of $\lambda_5$, 
$\lambda_5=1$ or $10^{-2}$ in the top panels while $\lambda_5=10^{-6}$ or $10^{-8}$ in the bottom panels.  
As we expect in Eq.~\eqref{eq:TR},  
the reheating temperature can be lowered by small $\lambda_5$ in a radiative seesaw model 
as compared with that in the Type-I seesaw model without taking large $R$. 
Thus, masses of right-handed neutrinos in a radiative seesaw model 
are not required to be very heavy for realizing successful reheating era leptogenesis. 
However, for $\lambda_5=10^{-9}$, all the parameter space is again excluded by perturbativity of 
the Yukawa coupling. 
As long as we use the reheating era leptogenesis scenario, 
the mass of the right-handed neutrino must be heavier than about $10^8$ GeV. 
A power law behavior of $T_R$ on $M_R$ is slightly different due to the function $F(M_R^2/M_\eta^2)$, 
and this behavior helps a little bit to extend allowed parameter space. 
As in the type-I case, the result without including the $\epsilon_2$ term is presented in the right panel of Figs.~\ref{fig:ma-NH} and \ref{fig:ma-IH}.
This effect is not large similarly to the type-I case.

%\begin{figure}[tb]
%\centering
% \includegraphics[width=7.1cm]{Ma-NH-HighMass-woe2.eps} 
% \includegraphics[width=7.1cm]{Ma-NH-LowMass-woe2.eps}
%\caption{
%The allowed parameter space of $T_R$ as a function of $M_R$ in Ma's radiative seesaw model. 
%The effect of the $\epsilon_2 $ term is omitted in Boltzmann equation.
%} 
%\label{fig:ma woe2}
%\end{figure}
%============================================================================

\section{Conclusion and Discussion}\label{sec4}

In this paper, we have extended the analysis of the letter article~\cite{Hamada:2015xva}.
we have applied the reheating era leptogenesis scenario 
to the various kinds of seesaw models for tiny neutrinos masses. 
It is shown that the reheating era leptogenesis can work not only in the Type-I (-\III) seesaw model 
but also the Ma's scotogenic seesaw model. 
In the seesaw models, the lepton number violation is related to the origin of neutrino masses, 
while in the above models there are sufficient freedoms to provide new CP violating phases. 
We have explicitly showed that CP violating phases really appear in the dimension-six term in the effective Lagrangian.
Compared with the letter article~\cite{Hamada:2015xva},
we have also examined new contributions to the reheating era leptogenesis, 
where the lepton number violating collision originated both from the inflaton decays. 
%
%This new contribution can become important for relatively small $T_R$ and large $M_R$ region. 
%
We have also studied several new constraints on the parameter space. 
Under these conditions, in each model, we have identified the allowed parameter space where the reheating era leptogenesis scenario works as a minimal alternative to thermal leptogenesis. 
We have found that the reheating temperature can be lower about $10^8$ GeV. 
An approximated analytic formula for a lower bound on $T_R$ is also presented. 
In the case of type-I seesaw model the lower bound on $T_R$ is proportional to $M_R^{-1/7}$, 
while a power law behavior of $T_R$ is slightly modified due to the function $F(M_R^2/M_\eta^2)$ 
in the scotogenic model.
This lower bound on $T_R$ puts the non-trivial constraint on inflation model, and is useful to discuss the unwanted relics/dark matter production in the early universe, see e.g. Refs.~\cite{gravitino,DM}.
The upper bound of $T_R$ is derived numerically, which is also new result of this paper.

In the Type-I seesaw model, the size of Yukawa coupling can be large 
by taking a large $R$, magnitude of the elements of a complex orthogonal matrix, 
if we allow a fine-tuning among model parameters. 
In the radiative seesaw models, the Yukawa coupling can be large enough for lowering $T_R$ 
with a new small parameter, e.g., $\lambda_5$ in the Ma's radiative seesaw model. 
The smallness of a new parameter can be easily explained by the naturalness argument 
relevant to the lepton number conservation and its breaking. 
Therefore, the reheating temperature can be lower generically in the radiative seesaw models. 
In this paper, we have concentrated on the models including right-handed neutrinos. 
However, this is not a necessary component in the reheating era leptogenesis scenario. 
It would be interesting to apply other variations of seesaw models.

%============================================================================

\subsection*{Acknowledgement}
Y.H. and D.Y. are supported by Japan Society for the Promotion of Science (JSPS) Fellowships for Young Scientists.
K.T.'s work is supported in part by the MEXT Grant-in-Aid for Scientific Research on Innovative Areas No. 16H00868, the JSPS Grant-in-Aid for Young Scientists (B) No. 16K17697, and the Supporting Program for Interaction-based Initiative Team Studies (Kyoto University).

\appendix
\section{The Boltzmann equations}
In this Appendix, we clarify how we discriminate the high and low energy leptons in the text, and present the derivation of the Boltzmann equations~\eqref{Eq:Boltzmann1}, \eqref{Eq:Boltzmann2} and \eqref{Eq:Boltzmann3}.

\begin{figure}[tb]
\centering
 \includegraphics[width=9.1cm]{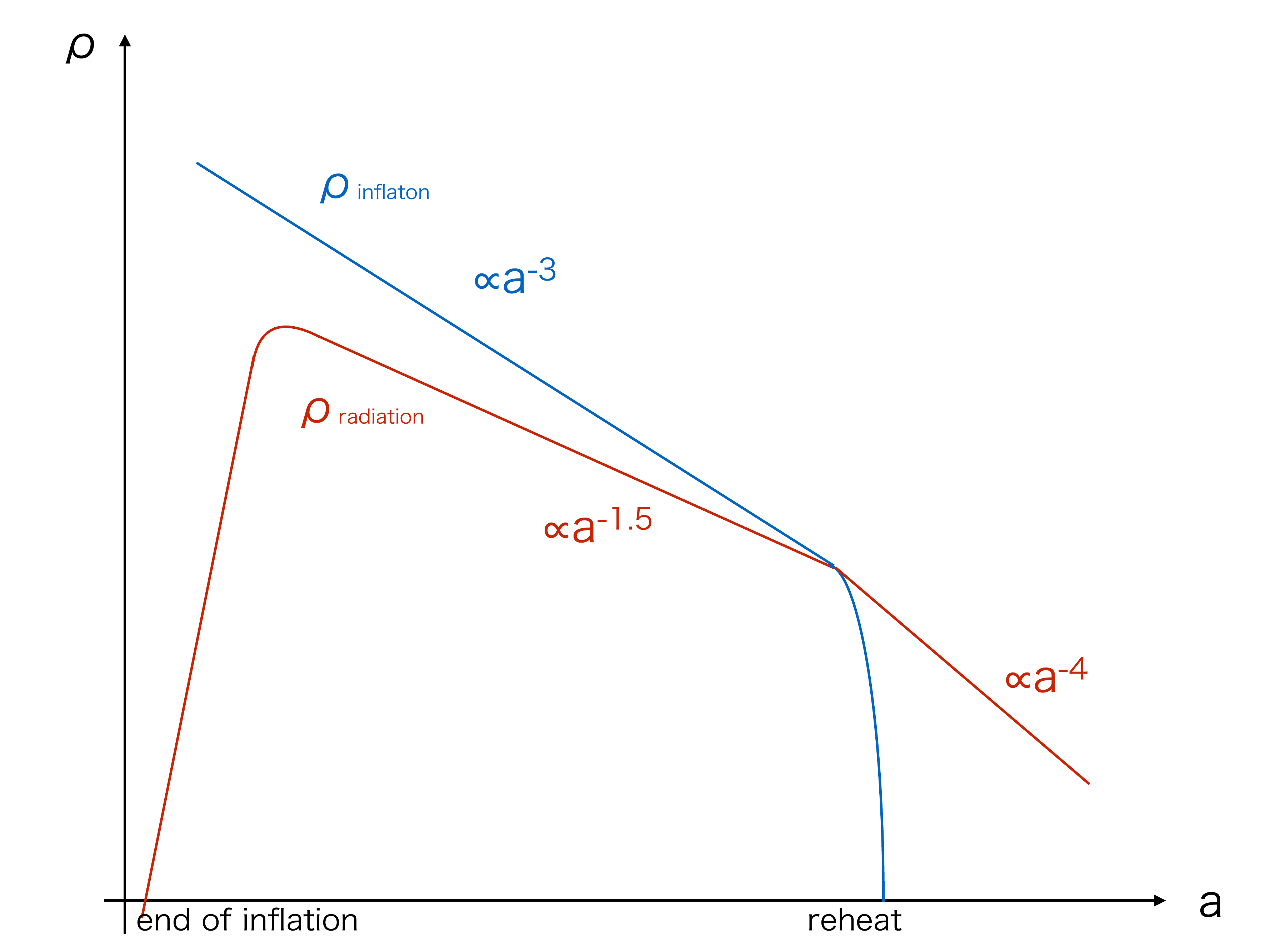} 
\caption{
A schematic picture for the energy densities of inflaton $\rho_{\text{inflaton}}$ and 
of radiation $\rho_{\text{radiation}}$ during the reheating process.
The horizontal axis is the scale factor of the universe $a$.
} 
\label{fig:reheating}
\end{figure}

Before going into details, let us explain the schematic picture of our scenario during the reheating.
We focus on the perturbative reheating scenario, which is one of the typical scenario of the reheating process, 
see, e.g., chapter 8 of Ref.~\cite{Kolb:1990vq} and Fig,~\ref{fig:reheating}.
In this scenario, after the end of inflation, the inflaton oscillation era starts.
In this era, an inflaton continues to decay until the end of reheating, and there exists the radiation component in addition to the inflaton energy density.
As long as thermalization rate is larger than Hubble rate, we can treat this radiation as thermal plasma.
Then, at around the completion of reheating, there are two populations of leptons.
One is generated by inflaton decay and the other is in thermal bath.
The interaction among them leads to the generation of lepton asymmetry of the universe.

Under the assumption that the universe is homogeneous and isotropic, the distribution function $f_{\ell_i}^{}$ for leptons is only the function of time $t$ and the absolute value of the three momentum $p=|\vec{p}|$. 
The Boltzmann equation is given by%\footnote{
%\red{
%As noted in the paragraph above, at each time $t_1$, there are two population of leptons, thermal and non-thermal ones.
%If we think the inflaton decay stops at some point, these two component thermalizes after the time $\sim\Gamma_\text{brems}^{-1}$ while the nonzero asymmetry is generated during this process.
%The resultant thermal distribution is $f=f_\text{th}$ appearing in Eq.(29).
%On the other hand, if the source term is present, there exists two population of leptons again after the time $\sim\Gamma_\text{brems}^{-1}$.
%Here the non-thermal population is leptons which appear during $t_1<t<t_1+\Gamma_\text{brems}^{-1}$ by inflaton decay, and thermal ones include leptons generated by inflaton decay for $t<t_1$.
%}
%}
%
\begin{align}\label{Eq:kinetic equation}
&\partial_t f_{\ell_i}(p,t)
-
Hp\,\partial_p f_{\ell_i}(p,t)
\nonumber\\
&=
{\Gamma_\text{inf}\,\rho_\text{inf}^{}\over M_\text{inf}}\, \mathcal{B}_i\, g(p) 
-
\left\{
f_{\ell_i}(p,t)-f_{\ell_i,\text{th}}(p,t)%\red{\left({\int dp'f_{\ell_i}(p',t)\over \int dp'f_{\ell_i,\text{th}}(p',t)}\right)}
\right\} \int (4\pi)q^2dq\, f_R^{}(q,t)\sigma_\text{brems}
,
%\nonumber\\
%&\dot{T}_{\ell_i}
%=
\end{align}
where $g(p)$ is the distribution function of leptons from the inflaton decay, $f_R^{}$ is the distribution for SM particles, and $f_{\ell_i,\text{th}}^{}$ is the thermal distribution function.
The normalization of $g(p)$ is $\int (4\pi)p^2dp\, g(p)=1$, and $\int (4\pi)p^2dp\, f_{\ell_i}(p,t)$ corresponds to the number density of lepton.
The left-hand side describes the time evolution of the distribution function with 
the expansion of the universe while the right-hand side does the collision terms.
Here, we only consider the following two processes; 
one is the decay of inflaton, and the other is thermalization whose bottleneck process is the bremsstrahlung 
with SM particles.
Since the thermalization process is dominated by the exchange of soft gauge bosons~\cite{Harigaya:2013vwa}, 
$\sigma_\text{brems}$ can be treated as a constant in the integral. 
%
%As for the second term in the right hand side,we introduce $({\int dp'f_{\ell_i}(p',t)/\int dp'f_{\ell_i,\text{th}}(p',t)})$ in such a way that the second term vanishes if we integrate over $p$.
%

%Eq.~\eqref{Eq:kinetic equation} shows that, in the absence of source term(first term) and expansion term, the distribution function 
The temperature of $f_{\ell_i,\text{th}}$ is determined by the requirement of the conservation of the energy density at fixed time,\footnote{
$f_\text{th}$ in Eq.~\eqref{Eq:kinetic equation} does not exactly equal to the thermal component discussed in paragraph above Eq.(29), although they are numerically similar. 
If we think that inflaton decay stops at some time, two components of leptons thermalize after the time $\sim\Gamma_\text{brems}^{-1}$. The resultant thermal distribution is $f=f_\text{th}$ appearing in Eq.(29).
In fact, in the absence of the source term, $f=f_\text{th}$ should be the solution of Boltzmann equation corresponding to the thermal equilibrium, and temperature is determined by taking into account all energy density.
This is why the total energy conservation is required in Eq.(30).
}
\begin{align}\label{Eq:energy conservation}
\sum_{k\neq i\text{-th lepton}}\int dp \,(4\pi)p^3 f_k
+
\int dp \,(4\pi)p^3 f_{\ell_i}
=
\sum_{k\neq i\text{-th lepton}}\int dp\, (4\pi)p^3 f_{k,\text{th}}
+
\int dp \,(4\pi)p^3 f_{\ell_i,\text{th}}
,
\end{align}
where $k$ is the label of SM particles except for $i$-th lepton.

In the following discussions, we neglect the second term in the left-hand side in Eq.~\eqref{Eq:kinetic equation}, 
because we are interested in the generation of the lepton asymmetry during the thermalization process, 
and the typical time scale of the thermalization is much faster than the Hubble time.
%
%We note that other processes such as pair production/annihillation is also be neglected because these rates are smaller than the thermalization rate.

%\red{Since (temperature with high energy lepton)\ll (without high energy lepton), Eq.(31) is still valid. At the era of the reheating completion, (temperature with high energy lepton)\simeq (without high energy lepton),}
We introduce pivot momentum $p_0^{}$ which is smaller than $M_\text{inf}$ and larger than $T_R$, 
and make an assumption of 
%
%\green{(We may need a justification of this inequality)}
%
\begin{align}\label{Eq:Boltzmann suppression}
&
\int_{p_0}^\infty (4\pi)p^2dp\,f_{\ell,\text{th}}(p,t)\ll \int_0^{p_0}(4\pi)p^2dp\, f_{\ell,\text{th}}(p,t),
&&
\int_{p_0}^\infty (4\pi)p^3dp\,f_{\ell,\text{th}}(p,t)\ll \int_0^{p_0}(4\pi)p^3dp\, f_{\ell,\text{th}}(p,t),
\end{align}
which can be justified in the case of $T_R\ll M_\text{inf}$ thanks to the Boltzmann suppression factor.
In fact, we are interested in the permitter region where $T_R<M_R<M_\text{inf}$.
The number densities for high and low energy leptons are defined as 
\begin{align}
&
\int_{p_0}^\infty (4\pi)p^2dp\,f_\ell(p,t)=:n_{\ell_i},
&
\int_0^{p_0} (4\pi)p^2dp\,f_\ell(p,t)=:n_{T_i}^{}.
\end{align}
Then, from Eq.~\eqref{Eq:kinetic equation}, we obtain 
\begin{align}\label{Eq:Boltzmann from kinetic}
\dot{n}_{\ell_i}
&=
{\Gamma_\text{inf}\rho_\text{inf}\over M_\text{inf}} \mathcal{B}_i
-
n_{\ell_i}\Gamma_\text{brems},
\\
\dot{\rho}_{T_i}^{}
&=
-(\rho_{T_i}-\rho_{\ell_i,\text{th}})\Gamma_\text{brems},
\label{Eq:rho_T}
\end{align}
where 
\begin{align}
&
\Gamma_\text{brems}:=\int (4\pi)q^2dq\, f_R^{}(q,t)\sigma_\text{brems},
&&
\rho_{T_i}:=\int_{0}^{p_0} (4\pi)p^3dp\,f_\ell(p,t),
&&
\rho_{\ell_i,\text{th}}:=\int_{0}^{p_0} (4\pi)p^3dp\,f_{\ell_i,\text{th}}(p,t)
\end{align}
Since the typical momentum of leptons from inflaton decay is $\mathcal{O}(M_\text{inf})$, 
we then expect $\int^\infty_{p_0}(4\pi)p^2 dp\, g(p)\simeq 1$ and $\int^{p_0}_0(4\pi)p^2 dp\, g(p)\simeq 0$. 
For $\Gamma_\text{brems}\gg H$, Eq.~\eqref{Eq:Boltzmann from kinetic} agrees with 
the Boltzmann equation \eqref{Eq:Boltzmann3} given in the text. 

Let us move on Eq.~\eqref{Eq:rho_T}.
Utilizing Eq.~\eqref{Eq:energy conservation}, this becomes
\begin{align}\label{Eq:rho_T2}
\dot{\rho}_{T_i}^{}
&=
\rho_{\ell_i}\Gamma_\text{brems}
+\sum_{k}(\rho_{k}-\rho_{k,\text{th}})\Gamma_\text{brems}.
\end{align}
Here $\rho_k:=\int_{0}^{\infty} (4\pi)p^3dp\,f_{k}(p,t), \rho_{k,\text{th}}:=\int_{0}^{\infty} (4\pi)p^3dp\,f_{k,\text{th}}(p,t)$, and we take $\int_0^\infty(4\pi)p^3dp f_{\ell,\text{th}} \simeq \int_0^{p_0}(4\pi)p^3dp f_{\ell,\text{th}}$ as in Eq.~\eqref{Eq:Boltzmann suppression}.
We notice that, with this approximation, Eq.~\eqref{Eq:energy conservation} becomes
\begin{align}
\sum_{k\neq i\text{-th lepton}}\rho_k
+
\left(\rho_{T_i}+\rho_{\ell_i}\right)
&=
\sum_{k\neq i\text{-th lepton}} \rho_{k,\text{th}}
+
\rho_{\ell_i,\text{th}}
,\nonumber\\
\Longrightarrow
-\left(\rho_{T_i}-\rho_{\ell_i,\text{th}}\right)
&=
\sum_{k\neq i\text{-th lepton}}(\rho_k- \rho_{k,\text{th}})
+\rho_{\ell_i}.
\end{align}
The equation like Eq.~\eqref{Eq:rho_T} also holds for other SM species:\footnote{
Regarding particles other than leptons, we do not distinguish high energy and low energy ones.
}
\begin{align}\label{Eq:SM particle}
\sum_k\dot{\rho}_{k}^{}
&=
{\Gamma_\text{inf}\rho_\text{inf}} (1-\mathcal{B}_i)-\sum_k(\rho_{k}-\rho_{k,\text{th}})\Gamma_\text{brems}.
\end{align}
Here we have used the fact that the typical energy of decay product of the inflation is $M_\text{inf}$, namely, $\int^\infty_{0}(4\pi)p^3 dp\, g(p)\simeq M_\text{inf}$. 
By combing Eqs.~\eqref{Eq:rho_T2} and \eqref{Eq:SM particle},  it is found that
\begin{align}
\dot{\rho}_{T_i}^{}+\sum_k\dot{\rho}_{k}^{}
&=
{\Gamma_\text{inf}\rho_\text{inf}} (1-\mathcal{B}_i)
+
\rho_{\ell_i}\Gamma_\text{brems},
\end{align}
which corresponds to \eqref{Eq:Boltzmann1} in the text.

Similarly, we can easily reproduce the Boltzmann equation for the lepton asymmetry.
We denote the distribution function for lepton asymmetry by $f_L(p,t)$, whose evolution is governed by
\begin{align}
&\partial_t f_{L}(p,t)
-
Hp\, \partial_p f_{L}^{}(p,t)
\nonumber\\
&=
-
\sum_i\int (4\pi)q_1^2dq_1(4\pi)q_2^2dq_2\,f_{\ell_i}(q_1,t)f_{\ell_i}(q_2,t)\sigma_{\cancel{L}}^{} \epsilon_i(q_1,q_2)
\left(\delta(q_1-p)+\delta(q_2-p)\right)
\nonumber\\
&-
f_{L}^{}(p,t)\int (4\pi)q^2dq\, \sigma_\text{wash} f_R^{}(q,t),
\end{align}
where $\sigma_{\cancel{L}}^{}$ is the cross section for the lepton number violating scattering, 
and $\sigma_\text{wash}$ is that of the washout process. 
The first and second terms in the right-hand side represent the lepton number production by the scattering 
and the washout effect, respectively.\footnote{As for the first term, we only take into account $L_iL_i \to \Phi\Phi$ 
(and $\bar{L}_i\bar{L}_i \to \bar{\Phi}\bar{\Phi}$) process. 
The other process such as $L_i\bar{\Phi} \to \bar{L}_i\Phi$ would give a similar contribution. 
We here omit the Pauli blocking effect and a stimulating emission factor.
}
Note that $\epsilon_i$ is proportional to the center of mass energy of the scattering, $\epsilon_i\propto q_1q_2$~\cite{Hamada:2015xva}.

As in the previous case, we integrate over $p$, and divide the momentum integral into two parts, and then get
\begin{align}
&\sum_i\int_0^\infty (4\pi)p^2dp\int_0^\infty (4\pi)q_2^2dq_2\,f_{\ell_i}(p,t)f_{\ell_i}(q_2,t)\sigma_{\cancel{L}} \epsilon_i(p,q_2)
\nonumber\\
&=
\sum_i
\bigg[
\int_{p_0}^\infty (4\pi)p^2dp\int_{p_0}^\infty (4\pi)q_2^2dq_2\,f_{\ell_i}(p,t)f_{\ell_i}(q_2,t)\sigma_{\cancel{L}} \epsilon_i(p,q_2)
\nonumber\\
&+
2\int_{p_0}^\infty (4\pi)p^2dp\int_0^{p_0}  (4\pi)q_2^2dq_2\,f_{\ell_i}(p,t)f_{\ell_i}(q_2,t)\sigma_{\cancel{L}} \epsilon_i(p,q_2)
\nonumber\\
&+
\int_0^{p_0} (4\pi)p^2dp\int_0^{p_0} (4\pi)q_2^2dq_2\,f_{\ell_i}(p,t)f_{\ell_i}(q_2,t)\sigma_{\cancel{L}} \epsilon_i(p,q_2)
\bigg]
\nonumber\\
&\simeq
\sum_i
\bigg[
n_{\ell_i} \Gamma_{2\cancel{L}} \epsilon_i\left({M_\text{inf}\over2},{M_\text{inf}\over2}\right)
+
2n_{\ell_i}\Gamma_{\cancel{L}} \epsilon_i\left({M_\text{inf}\over2},3T\right)
\bigg]
%\nonumber\\
%&+
%\int_0^{p_0} dq_1\int_0^{p_0} dq_2f_{\ell_i}(q_1,t)f_{\ell_i}(q_2,t)\sigma_{\cancel{L}} \epsilon_i(q_1,q_2)
\end{align}
In the last step, we have made an approximation.
From Eq.~\eqref{Eq:kinetic equation}, we see that, if the cosmic expansion is neglected and 
the initial condition at $t=t_\text{initial}$ (end of the inflation) is $f_{\ell_i}(p,t_\text{initial})=0$, 
the distribution function of leptons is peaked at around $\mathcal{O}(M_\text{inf})$ and $\mathcal{O}(T)$. 
Moreover, because the evolution equation is
\begin{align}
%&\partial_t f_{\ell_i}(p,t)
%\simeq 
%{\Gamma_\text{inf}\,\rho_\text{inf}^{}\over M_\text{inf}}\,g(p)
%-
%f_{\ell_i}(p,t)
% \int (4\pi)q^2dq\, f_R^{}(q,t)\sigma_\text{brems}
%, \quad p\simeq M_\text{inf},
%\nonumber\\
&\partial_t f_{\ell_i}(p,t)
\simeq 
-
\left\{
f_{\ell_i}(p,t)-f_{\ell_i,\text{th}}(p,t)
\right\} \int (4\pi)q^2dq\, f_R^{}(q,t)\sigma_\text{brems}
%, \quad p< M_\text{inf},
\end{align}
for $p< M_\text{inf}$, one can see that the distribution function of leptons with momentum $p< M_\text{inf}$ is proportional to thermal one together with $f_{\ell_i}(p,t_\text{initial})=0$.
Therefore, we can replace the momenta which appear in $\epsilon_i$ by their typical values. 
As a concrete value, we put $M_\text{inf}/2$ and $3T$, which are typical scales of the inflaton decay   
and the thermal bath, respectively.
We omit the last term in the second line because the distribution function in the term is close to 
the thermal distribution, which does not contribute the lepton asymmetry~\cite{Hamada:2015xva}.
%
%
%As a result, we obtain
%%
%\begin{align}
%&\int_0^\infty dq_1\int_0^\infty dq_2f_{\ell_i}(q_1,t)f_{\ell_i}(q_2,t)\sigma_{\cancel{L}} \epsilon_i(q_1,q_2)
%\nonumber\\
%&\simeq
%n_{\ell_i} \Gamma_{2\cancel{L}} \epsilon_i\left({M_\text{inf}\over2},{M_\text{inf}\over2}\right)
%+
%2n_{\ell_i}\Gamma_{\cancel{L}} \epsilon_i\left({M_\text{inf}\over2},3T\right).
%\end{align}
%
%
We use the following notations for the equations given in the text; 
\begin{align}
&
\int (4\pi)q^2dq\, \sigma_\text{wash} f_R(q,t)=:\Gamma_\text{wash},
&&
\int_0^{p_0} (4\pi)q^2dq\, \sigma_{\cancel{L}} f_{\ell_i}(q,t)=:\Gamma_{\cancel{L}_i},
&&
\int_{p_0}^\infty (4\pi)q^2dq\, \sigma_{\cancel{L}} f_{\ell_i}(q,t)=:\Gamma_{2\cancel{L}_i}.
&
\end{align}
Note that $\sigma_{\cancel{L}}^{}$ is constant as long as the center of mass energy is lower than 
the mass of right-handed neutrinos.

By combing these above arguments, we arrive at
\begin{align}
\dot{n} _{L}(p,t)
=&
2\left(
2n_{\ell_i}\Gamma_{\cancel{L}} \epsilon_i\left({M_\text{inf}\over2},3T\right)
+
n_{\ell_i} \Gamma_{2\cancel{L}} \epsilon_i\left({M_\text{inf}\over2},{M_\text{inf}\over2}\right)
\right)
-
n_{L}\Gamma_\text{wash},
\end{align}
which reproduce the Boltzmann equation~\eqref{Eq:Boltzmann2} for $\Gamma_\text{brems}\gg H$.

\end{document}